\documentclass[iop,superscriptaddress]{emulateapj}

\usepackage[caption=false]{subfig}
\usepackage{epsfig}
\usepackage{verbatim}
\usepackage{graphicx}

\usepackage{xcolor}
\definecolor{darkgreen}{rgb}{0.0,0.5,0.0}
\usepackage[colorlinks,citecolor=darkgreen]{hyperref} % Works with PDFLaTeX; use for arXiv

\newcommand{\Eph}{{\epsilon}}
\newcommand{\Ee}{{E}}

\newcommand{\gama}{{$\gamma$}}
\newcommand{\kpc}{\mbox{ kpc}}
\newcommand{\pc}{\mbox{ pc}}
\newcommand{\Mpc}{\mbox{ Mpc}}
\newcommand{\se}{\mbox{ s}}
\newcommand{\cm}{\mbox{ cm}}
\newcommand{\km}{\mbox{ km}}
\newcommand{\erg}{\mbox{ erg}}

\newcommand{\eV}{\mbox{ eV}}
\newcommand{\keV}{\mbox{ keV}}
\newcommand{\MeV}{\mbox{ MeV}}
\newcommand{\GeV}{\mbox{ GeV}}
\newcommand{\TeV}{\mbox{ TeV}}

\newcommand{\GHz}{\mbox{ GHz}}
\newcommand{\K}{\mbox{ K}}
\newcommand{\mK}{\mbox{ mK}}
\newcommand{\muG}{\mbox{ $\mu$G}}
\newcommand{\sr}{\mbox{ ster}}
\newcommand{\dgr}{{^\circ}}

\newcommand{\vect}[1]{\mathbf{#1}}
\newcommand{\ie}{\emph{i.e.} }
\newcommand{\eg}{\emph{e.g.,} }
\newcommand{\fin}{\mbox{ .}}
\newcommand{\coma}{\mbox{ ,}}

\newcommand{\Non}{N_{\mbox{\scriptsize{on}}}}
\newcommand{\Noff}{N_{\mbox{\scriptsize{off}}}}
\newcommand{\Aon}{A_{\mbox{\scriptsize{on}}}}
\newcommand{\Aoff}{A_{\mbox{\scriptsize{off}}}}
\newcommand{\non}{n_{\mbox{\scriptsize{on}}}}
\newcommand{\noff}{n_{\mbox{\scriptsize{off}}}}
\newcommand{\thetaon}{\theta_{\mbox{\scriptsize{on}}}}
\newcommand{\thetar}{\theta_{r}}

\newcommand{\myzeta}{{\zeta}}
\newcommand{\myzetaT}{{\tilde{\zeta}}}
\newcommand{\mykappa}{{\delta}}
\newcommand{\mys}{{C}}
\newcommand{\phN}{{N}}
\newcommand{\pxN}{{\mathsf{n}}}
\newcommand{\Tcmb}{{T_{\mbox{\scriptsize cmb}}}}
\newcommand{\ucmb}{{u_{\mbox{\scriptsize cmb}}}}

\newcommand{\TSZ}{{T_{\mbox{\scriptsize sz}}}}
\newcommand{\Tsyn}{{T_{\mbox{\scriptsize syn}}}}
\newcommand{\myw}{{\mathsf{w}}}
\newcommand{\udot}{{\dot{u}}}
\newcommand{\mdot}{{\dot{m}}}
\newcommand{\mass}{{\bar{m}}}

\newcommand{\till}{{\mbox{--}}}

\newcommand{\mynewcommand}[2]{\ifdefined #1 \else \newcommand{#1}{#2} \fi}

\newcommand{\myNi}{\emph{(i)}\,}
\newcommand{\myNii}{\emph{(ii)}\,}
\newcommand{\myNiii}{\emph{(iii)}\,}
\newcommand{\myNiv}{\emph{(iv)}\,}

\newcommand{\mytwocolumn}{}
\newcommand{\myonecolumn}{}

\mynewcommand{\apj}{ApJ}     % Astrophysical Journal
\mynewcommand{\apjl}{ApJL}     % Astrophysical Journal
\mynewcommand{\apjs}{ApJS}    % Astrophysical Journal, Supplement
\mynewcommand{\aap}{A\&A}    % Astronomy and Astrophysics
\mynewcommand{\nat}{Nature}  % Nature
\newcommand{\MakeDouble}{0}

\defcitealias{VERITAS12_Coma}{V12}
\newcommand{\VER}{{\citetalias{VERITAS12_Coma}}}

\newcommand{\ApJMark}[1]{{#1}}
\newcommand{\Fermi}{{\emph{Fermi}} }
\begin{document}

\renewcommand*\thesection{\arabic{section}}

\if \MakeDouble 1
\setlength{\columnsep}{15pt}
\mytwocolumn
\fontsize{11pt}{8pt}\selectfont
\parindent 0.2in
\sloppy
\fi

\title{
Preliminary evidence for a virial shock around the Coma galaxy cluster
}

\author{
Uri Keshet\altaffilmark{1},
Doron Kushnir\altaffilmark{2},
Abraham Loeb\altaffilmark{3},
and
Eli Waxman\altaffilmark{4}
}

\email{ukeshet@bgu.ac.il}

\altaffiltext{1}{
Physics Department, Ben-Gurion University of the Negev, Be'er-Sheva 84105, Israel
}

\altaffiltext{2}{
Institute for Advanced Study, Einstein Drive, Princeton, New Jersey 08540, USA
}

\altaffiltext{3}{
Harvard-Smithsonian Center for Astrophysics, 60 Garden St., Cambridge, MA 02138, USA
}

\altaffiltext{4}{
Physics Faculty, Weizmann Institute of Science, POB 26, Rehovot, Israel
}

\begin{abstract}
Galaxy clusters, the largest gravitationally bound objects in the Universe, are thought to grow by accreting mass from their surroundings through large-scale virial shocks.
Due to electron acceleration in such a shock, it should appear as a $\gamma$-ray, hard X-ray, and radio ring, elongated towards the large-scale filaments feeding the cluster, coincident with a cutoff in the thermal Sunyaev-Zel'dovich (SZ) signal.
However, no such signature was found until now, and the very existence of cluster virial shocks has remained a theory.
We find preliminary evidence for a large, $\sim 5\Mpc$ minor axis $\gamma$-ray ring around the Coma cluster, elongated towards the large scale filament connecting Coma and Abell 1367\ApJMark{, detected at the nominal $2.7\sigma$ confidence level ($5.1\sigma$ using control signal simulations)}.
The $\gamma$-ray ring correlates both with a synchrotron signal and with the SZ cutoff, but not with Galactic tracers.
The $\gamma$-ray and radio signatures agree with analytic and numerical predictions, if the shock deposits $\sim 1\%$ of the thermal energy in relativistic electrons over a Hubble time, and $\sim 1\%$ in magnetic fields.
The implied inverse-Compton and synchrotron cumulative emission from similar shocks can significantly contribute to the diffuse extragalactic $\gamma$-ray and low frequency radio backgrounds.
Our results, if confirmed, reveal the prolate structure of the hot gas in Coma, the feeding pattern of the cluster, and properties of the surrounding large scale voids and filaments.
The anticipated detection of such shocks around other clusters would provide a powerful new cosmological probe.
\end{abstract}

\maketitle

%\sloppy

\section{Introduction}

In the hierarchical paradigm of large-scale structure (LSS) formation, galaxy clusters are the largest objects ever to virialize. With masses in excess of $10^{14}M_\odot$, they are located at the nodes of the cosmic web, where they accrete matter from the surrounding voids and through large-scale filaments. With their vast size, galaxy clusters resemble remote island universes, providing a powerful cosmological probe and a unique astrophysical laboratory.

The gas accreted by a cluster is thought to abruptly heat and slow down in a strong virial shock wave surrounding the cluster.
Such collisionless shocks should, by analogy with supernova remnant shocks, accelerate charged particles to $>\TeV$ energies, where they Compton-scatter cosmic microwave-background (CMB) photons up to the \gama-ray band \citep{LoebWaxman00, TotaniKitayama00, KeshetEtAl03}.
Consequently, one expects to find \gama-ray rings around clusters \citep{WaxmanLoeb00}, as indicated by cosmological simulations \citep{KeshetEtAl03, Miniati02}, which suggest an elliptic morphology elongated towards the large-scale filaments feeding the cluster \citep{KeshetEtAl03}.
Such rings are also expected in hard X-rays \citep{KushnirWaxman10}, and should coincide with a synchrotron radio ring \citep{WaxmanLoeb00,KeshetEtAl04,KeshetEtAl04_SKA} and with a cutoff in the thermal Sunyaev-Zel'dovich (SZ) signal \citep{KocsisEtAl05}.

However, no such virial shock signature has been detected until now, although a stacking analysis of EGRET data around a sample of 447 rich clusters did suggest a $3\sigma$ signal \citep{ScharfMukherjee02}.
Upper limit were imposed on the \gama-ray emission from clusters such as Coma \citep[][henceforth \VER]{SreekumarEtAl96, ReimerEtAl03, AckermannEtAl10, VERITAS12_Coma}\ApJMark{, including analyses of Coma based on \Fermi data \citep{AckermannEtAl10, AckermannEtAl14_GammaRayLimits}}, mostly focusing on the central parts of the cluster, well within the virial radius\ApJMark{; see discussion in \S\ref{subsec:PrevEstimates}}.
The very existence of cluster-scale virial shocks has thus remained unconfirmed.

\newsavebox{\myimage}
\begin{figure*}
  \centering
  \savebox{\myimage}{\makebox{\epsfxsize=8.5cm \epsfbox{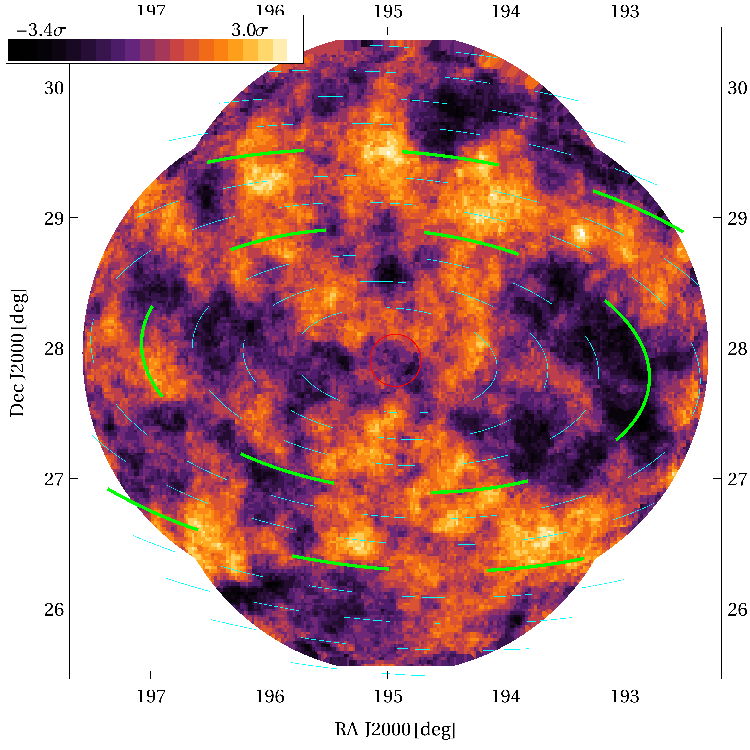}}}% Store largest image
  \subfloat[VERITAS significance map\label{fig:ComaVer}]{\usebox{\myimage}} \quad
  \subfloat[Simulated cluster in \gama-rays\label{fig:ComaSim}]{\raisebox{\dimexpr.5\ht\myimage-.48\height\relax}{\epsfxsize=9.2cm \epsfbox{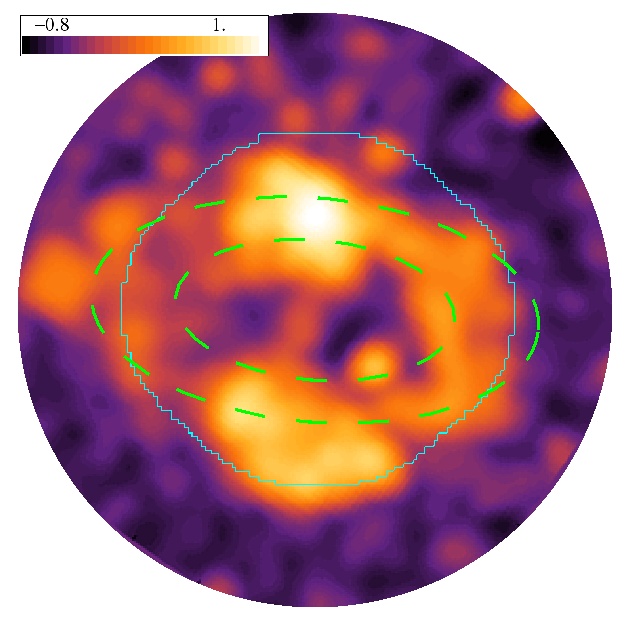}}}
   \caption{\label{fig:ComaVerSim}
{\bf Observed and simulated \gama-ray maps of Coma} (notice the different scales).\\
\emph{Left:} VERITAS $\gtrsim 220\GeV$ nominal significance map (\VER) of Coma for $\theta=0.2\dgr$ integration (illustrated by the red central circle). Elliptical bins are shown (thin dashed cyan contours) for $\Delta b=0.2\dgr$, $\myzeta\equiv a/b=2$ and $\phi=-5\dgr$.
The bins showing enhanced emission are highlighted (bounded by thick, long-dashed green curves).\\
\emph{Right:}
Simulated map of a Coma-like cluster from a $\Lambda$CDM simulation \citep{KeshetEtAl03}.
The largest in the $200\Mpc$ simulation box, this cluster has \citep{KeshetEtAl04} mass $M\simeq 10^{15}M_\odot$ and temperature $k_B T\simeq 8\keV$, like Coma.
The $8.5\dgr$ diameter image was convolved with a $\sim 0.23\dgr$ beam, comparable to the VERITAS map, and rotated such that the large-scale filament extends to the west.
Colorbar: $\log_{10}(J/10^{-8}\cm^{-2}\se^{-1}\sr^{-1})$ brightness above $220\GeV$ for acceleration efficiency $\xi_e\mdot=1\%$.
The regions corresponding to the VERITAS mosaic (solid cyan contour) and to the VERITAS ring (elliptic dashed contours) are similarly highlighted.\\
}
\end{figure*}

The Coma cluster (Abell 1656) is one of the richest nearby clusters. With mass $M\sim 10^{15}M_\odot$, temperature $k_B T\sim 8\keV$, and richness class 2, it lies only $\sim 100\Mpc$ away \citep{GavazziEtAl09}, at a redshift $z\simeq 0.023$.
The cluster lies near the north Galactic pole (latitude $\sim88\dgr$), in a sky patch remarkably devoid of Galactic foreground.
These considerations, and indications for a high accretion rate as discussed below, render Coma exceptionally suitable for the search for virial ring signatures.

The virial radius of Coma, $R\simeq R_{200}\simeq 2.3\Mpc$, corresponds to an angular radius $\psi\simeq \psi_{200}\simeq 1.3\dgr$.
Here, subscripts $200$ refer to an enclosed density $200$ times above the critical \ApJMark{density of the Universe}.
The cluster is somewhat elongated in the east--west direction, in coincidence with the western LSS filament \citep{WestEtAl95} that connects it with the cluster Abell 1367 (see Figure \ref{fig:SDSS}). There is X-ray \ApJMark{\citep{SimionescuEtAl13, UchidaEtAl16},} optical, weak lensing \ApJMark{\citep{OkabeEtAl10,OkabeEtAl14},} radio \citep{BrownRudnick11}, and SZ \citep{PlanckComa12} evidence that the cluster is accreting clumpy matter and experiencing weak shocks towards the filament well within the virial radius, at $\psi\sim0.5\dgr$ radii.

The VERITAS \v{C}erenkov array has produced a $d\sim 4.8\dgr$ diameter \gama-ray mosaic (\VER) of Coma, at energies $\Eph\gtrsim 220\GeV$.
We argue that the significance map (Figure \ref{fig:ComaVer})
shows evidence for extended \gama-ray emission away from the center, that appears as a ($\sim 0.6\dgr$) thick elliptical ring with semi-minor axis $b\simeq 1.3\dgr$, elongated along the east--west direction (best fit: $\phi\simeq-5\dgr$), with semi-major to semi-minor axes ratio $\myzeta\equiv a/b\gtrsim 3$.
The nominal significance of the signal is $S=2.7\sigma$, but accounting for the background removal indicates a $5.1\sigma$ significance (for a blind search in $b$, $\zeta$ and $\phi$).
This preliminary signal agrees with analytic and numerical predictions, and correlates with other tracers of the shock.

The paper is organized as follows.
We analyze the VERITAS mosaic in \S\ref{sec:VERITAS}, present additional evidence for extended \gama-ray emission in \S\ref{sec:AdditionalEvidence}, and show that other VERITAS fields show no such signal in \S\ref{sec:OtherFields}.
In \S\ref{sec:Simulation} we present the simulated VERITAS signature of a \ApJMark{Coma-like} virial shock.
The signal is shown to be unaffected by the Galactic foreground in \S\ref{sec:Galactic}, but positively correlated with expected radio signals in \S\ref{sec:Radio}.
We show in \S\ref{sec:InverseCompton} that the \gama-ray signal corresponds to an acceleration efficiency (over a Hubble time) of $\sim1\%$, and in \S\ref{sec:RadioTheory} that the radio correlations are consistent with $\sim1\%$ magnetization efficiency.
Finally, we summarize and discuss our results in \S\ref{sec:Discussion}. A simple $\beta$-model for emission from the virial shock is given in Appendix \S\ref{sec:BetaModel}.

We adopt a concordance flat $\Lambda$CDM model with Hubble constant $H=70\km\Mpc^{-1}$, a baryon fraction $f_b=17\%$, and a hydrogen mass fraction $\chi=0.76$.
The plasma is approximated as an ideal gas with an adiabatic index $\Gamma=5/3$ and a mean particle mass $\mass=0.59m_p$, where $m_p$ is the proton mass.

\section{VERITAS \gama-ray ring around Coma}
\label{sec:VERITAS}

The VERITAS collaboration has presented (\VER) a $d\sim 4.8\dgr$ diameter \gama-ray mosaic of Coma, at energies $\gtrsim 220\GeV$. The significance map (Figure \ref{fig:ComaVer}) suggests some extended \gama-ray emission away from the center.
The \gama-ray structure appears as an elliptical ring, elongated along the east-west direction, or as two parallel east-west filaments lying symmetrically both north and south of the cluster.
The spectrum of the VERITAS feature is probably flat, as a $p=2.4$ photon spectral index was used to optimize the gamma-hadron separation cuts (\VER).
In \S\ref{sec:InverseCompton} we show that a flat spectrum with index $p=2$ (equal energy per logarithmic energy interval, as expected in the strong virial shock) is consistent with observations at both lower and higher energies.

\subsection{Foreground removal: ring model}

Due to the strong foreground, the significance $s_j$ at a given direction $j$ in the VERITAS map is estimated by comparing the numbers of events arriving from the target vicinity (the so-called on region), $\Non$, and away from it (the off region), $\Noff$.
These counts are then weighted by the detector acceptance, which varies as a function of the angular distance from the detector pointing.
Defining the acceptance integrated over the on and off regions, $\Aon$ and $\Aoff$, and their ratio $\bar{\alpha}\equiv \Aon/\Aoff$, a positive (negative) signal corresponds to a positive (negative) value of $(\Non-\bar{\alpha}\Noff)$.

The {\VER} data were collected in $\myw=0.5\dgr$ wobble mode \citep{DanielEtAl08}, and analyzed using the ring background model \citep{BergeEtAl07} with a $\thetaon=0.2\dgr$ integration radius for the on region. The significance $s_j$ was computed from $\Non$, $\Noff$ and $\bar{\alpha}$  based on the likelihood ratio method \citep[][equation 17]{LiMa83}.
The background can be estimated from a $\theta^2$ analysis (\VER), indicating $\phN_{0.2}\simeq 1200$ counts above $220\GeV$ in the central $\theta=0.2\dgr$ beam.
The background per pixel is approximately $\phN_B=\phN_{0.2}/\pxN_{0.2}$, where $\pxN_{0.2}$ is the number of pixels in the beam.

We wish to compute the significance of an extended feature, in particular an elliptic ring around Coma.
For each pixel $j$, we estimate the numbers of counts per pixel in both on and off regions, $\non$ and $\noff$, normalized such that the ratio of integrated acceptance over the two regions is unity. Integrating the corresponding excess and total counts over an extended source then yields its overall significance \citep{LiMa83},
\begin{equation}
S = \frac{\sum_j \left(\non-\noff\right)}{\left[ \sum_j \left(\non+\noff\right) \right]^{1/2} } \, .
\end{equation}

To find $\non$ and $\noff$, we combine the significance map with the simulated acceptance profiles of VERITAS \citep[linearly interpolated from Figure 4 of][]{MaierEtAl08} for each pointing, adopt the mean ring radius $\thetar=0.5\dgr$ and solid angle ratio $\Omega_{\mbox{\scriptsize{off}}}/\Omega_{\mbox{\scriptsize{on}}}=7$ typical of the ring model \citep{BergeEtAl07}, make the approximation $\bar{\alpha}\Noff\simeq \Aon \phN_B$, and derive $\Non$ from $s_j$ using the likelihood ratio method.

\subsection{East-west thick elongated ring: \ApJMark{nominal} $2.7\sigma$ }
\label{sec:NominalSignificance}

In order to quantify the \gama-ray structure and assess its significance, we fit the data with a thick, elliptical or filamentary virial ring model.
The center of the virial ring is chosen as the cluster's X-ray peak in the \emph{ROSAT} All Sky Survey \citep[RASS;][]{SnowdenEtAl97}.
The semi-major axis $a$ is taken along the east-west direction, as inferred from the \emph{ROSAT} map and from the orientation of the SDSS galaxy filament.
The thickness of the ring is fixed at $\Delta b=0.6\dgr$ along the semi-minor axis, because a thicker ring would exceed the maximal $\thetar+\Delta\thetar=0.64\dgr$ limit imposed by the ring-model background subtraction.

A thick, east--west filamentary or very elongated ring model with a median semi-minor axis given by the virial radius, $b=\psi_{200}\simeq 1.3\dgr$, is thus well-defined, and has no free parameters.
Such an extended structure is seen at an $S=2.7\sigma$ nominal significance level; no trial factors ('look elsewhere' corrections) are necessary.

\begin{figure}[h]
\centerline{\epsfxsize=9cm \epsfbox{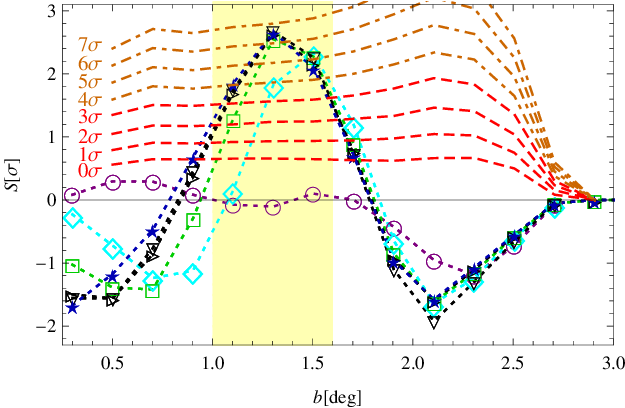}
}
\caption{\label{fig:ComaVerBin}
{\bf Significance of the VERITAS ring.\\}
The VERITAS signal (Figure \ref{fig:ComaVer}) fit with $\Delta b=0.6\dgr$ thick, east-west elongated rings (symbols; connecting lines are guides to the eye) of major-to-minor axes ratios $\myzeta\equiv a/b=1$ (circles), $2$ (diamonds), $3$ (squares), $5$ (triangles, pointed right for $\phi=0$ and down for $\phi=-5\dgr$) and $1000$ (stars).
The yellow band corresponds to the bright extended ring marked in Figure \ref{fig:ComaVer}.
For comparison, the distribution of the maximal significance rings in mock VERITAS maps is measured (dashed lines) and extrapolated (linearly; dot-dashed).\\
}
\end{figure}

\begin{figure}[h]
\centerline{
\epsfxsize=7cm \epsfbox{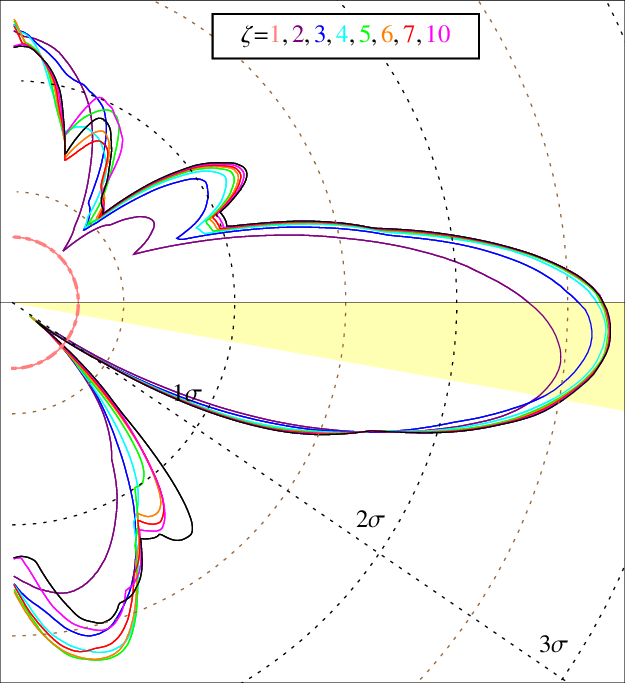}
}
\caption{\label{fig:ComaVerTilt}
\if \MakeDouble 1
\fontsize{10pt}{8pt}\selectfont
\fi
{\bf Orientation of the VERITAS ring.\\}
Nominal significance $S$ (solid) curves are shown for $\Delta b=0.6\dgr$ rings of various ellipticity $\myzeta$ (legend) and orientation $\phi$.
The yellow sector corresponds to the $-10\dgr<\phi<0\dgr$ orientation of the LSS filament.
The secondary peak towards $\phi\simeq 70\dgr$ with $b\sim 0.5\dgr$ is related to the primary peak: the semi-major axis of the small ellipse solution roughly coincides with the semi-minor axis of the large ellipse solution.\\
}
\end{figure}

A less elongated, east--west ring is defined by $b$ and by the elongation ratio $\myzeta\equiv a/b$.
For sufficient elongation, $\myzeta\gtrsim 3$, the results (Figures \ref{fig:ComaVerBin} and \ref{fig:ComaVerTilt}) are found to depend weakly upon $\myzeta$.
Scanning for the most significant structure as a function of $b$ then indeed indicates an extended structure around $b\simeq 1.3\dgr$.
Here, the $S=2.7 \sigma$ nominal significance level should be corrected for trial factors in $b$.
However, the nominal significance remains above $2\sigma$ over the plausible range of $1.1\dgr<b<1.5\dgr$, and remains positive even for scales as small as $\psi_{500}$ or as large as $\psi_{100}$.
Similarly, fixing $b=\psi_{200}$ while varying the elongation gives $S>1.8\sigma$ for the entire plausible, $\zeta>2$ range.

The SDSS (Figure \ref{fig:SDSS}) and \emph{ROSAT} maps suggest that the LSS filament and the semi-major axis of the gas distribution are slightly tilted towards the southwest, at an angle $-10\dgr\lesssim\phi\lesssim 0\dgr$.
Indeed, the significance of the \gama-ray structure is maximized when thus tilted (Figure \ref{fig:ComaVerTilt});
the best fit is obtained for $\myzeta\gtrsim 4$, $\phi\simeq -5\dgr$.
One way to see that the signal is significant even when accounting for trial factors in $\phi$ is by comparison to other VERITAS observations; this is done in \S\ref{sec:OtherFields}.
First, we compute the significance of the structure while accounting for trials in all parameters using Monte-Carlo simulations; for this we must address the details of the observation mode and data preparation.

\subsection{Monte Carlo significance estimate: $5.1\sigma$}
\label{sec:TrueSignificance}

Any extended, $\gtrsim \thetar=0.5\dgr$ feature would be partly erased by the ring background model.
Hence, $S$ can substantially underestimate the true significance of an extended structure such as a thick ring.
A more reliable estimate of the signal significance may be obtained by applying the same elliptical ring template presented in \S\ref{sec:NominalSignificance} to mock significance maps, based on random photon noise but no signal, and analyzing the resulting $S$ distribution.
We prepare $>10^4$ such mock maps, by generating Poisson noise flux at the observed background level, filtering the flux through the $0.5$ wobble mode acceptance of VERITAS, integrating over the same $\theta=0.2\dgr$ beam using the ring background model, and computing the per-pixel significance using the likelihood ratio method.

\begin{figure*}[t]
\centerline{\epsfxsize=10cm \epsfbox{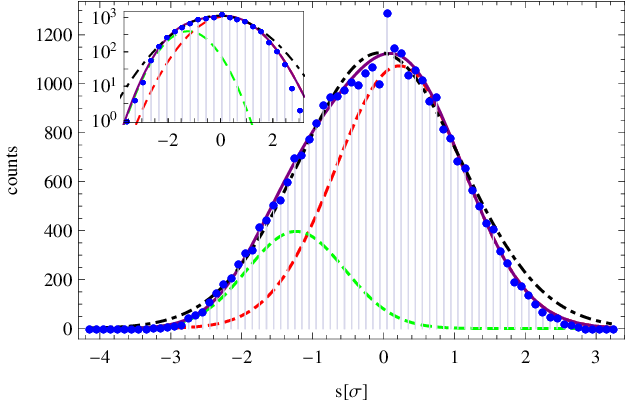}}
\caption{\label{fig:VER_Hist}
\if \MakeDouble 1
\fontsize{10pt}{8pt}\selectfont
\fi
{\bf Histogram of pixel significance in the VERITAS map.\\}
A single Gaussian function (dot-dashed curve) provides a marginal fit to the data.
A much better fit is provided by the sum (solid) of two Gaussian functions (dashed); see text for function parameters.
Inset: logarithmic histogram.\\
}
\end{figure*}

We may now obtain a more reliable estimate of the significance of the extended VERITAS ring, by searching for such a feature in the mock maps, using the exact same method applied to the true VERITAS mosaic.
We still assume that the mock ring is thick ($\Delta b=0.6\dgr$, fixed by the ring background model), but make no assumptions regarding its ellipticity $\myzeta$ or orientation $\phi$.
Rather, for each $b$, we identify the most significant ring, of arbitrary ellipticity (scanning $2\leq\myzeta\leq7$ with a $0.5$ step) and orientation (scanning $-90\dgr\leq\phi\leq+90\dgr$ with a $10\dgr$ step).
The resulting distribution $\Sigma(S,b)$ of highest-significance rings is shown as the nearly horizontal curves in Figure \ref{fig:ComaVerBin}.
Keeping either $\zeta$ or $\phi$ fixed, while varying the other, yields qualitatively similar distributions.
The enhanced $\Sigma$ above $b\simeq 2\dgr$ may be partly due to edge effects.

None of the mock maps shows a ring as significant as the observed \ApJMark{nominal} $S=2.7\sigma$ feature.
Due to the limited mock sample size, we can compute confidence levels only up to $\sim4\sigma$, shown as dashed lines in Figure \ref{fig:ComaVerBin}; this places a lower, $4\sigma$ limit on the significance of the VERITAS signal.
(We use $\Sigma=1\sigma, 2\sigma, 3\sigma,\ldots$ to designate levels exceeded by $16\%, 2.3\%, 0.13\%, \ldots$ of the sample; henceforth.)
The accessible, $\Sigma\lesssim4\sigma$ mock sample is approximately linearly proportional to $S$, so an extrapolation to high significance is possible (dot-dashed lines in Figure \ref{fig:ComaVerBin}).
This assigns the VERITAS ring with a $5.1\sigma$ significance, if we allow for any $b<2\dgr$, $\zeta$ and $\phi$ \ApJMark{values}, and average the mock sample over all parameters.

Notice that if we were to constrain $b$ in this analysis to values near the expected, virial radius, or constrain $\phi$ to orientations coincident with the LSS filament, then the Monte-Carlo based estimate for the signal significance would become even higher.

\section{Further evidence for extended emission}
\label{sec:AdditionalEvidence}

Consider the correlation between \gama-ray pixels $s_j$ induced by the ring model background subtraction.
Due to this non-local procedure, the significance of a pixel is anti-correlated with the pixels within $\sim\thetar$ around it. One implication is that sources that extend over scales much larger than $\thetar$ are effectively erased by the background subtraction, except in an $\sim \thetar$ wide band in their periphery. In such a band, and in sources that extend $\lesssim \thetar$ in at least one direction, the signal is diminished but not completely erased.

\subsection{Extended feature with diminished $S$ and dark edges}

For example, the estimated, positive significance at the inner edge of a large linear source will be diminished by a factor of $\sim2$, and the signal at the outer edge will have an approximately equal but negative \ApJMark{nominal} significance.
(For simplicity, here we approximate the procedure as subtraction of the $\bar{\alpha}\Noff$ counts from the $\Non$ counts.)
Therefore, for sources that extend over $>\thetar$ in at least one direction, the actual significance may be $\gtrsim 2$ times higher than estimated from the ring model. This is consistent with the above estimates based on the mock sample analysis.

Another indication that the background subtraction has removed an extended signal is the negative, $\lesssim -3\sigma$ significance regions, lying just outside the bright feature we identify as an elliptical ring.
If we assume that the bright region is part of an extended, linear structure, this suggests that its local significance should be $\Delta s_j\gtrsim -(-3)/2 \simeq 1.5\sigma$ higher than it nominally appears, again consistent with the mock sample analysis.

A \gama-ray structure that extends over scales much greater than $\thetar$ would be largely erased by the ring background subtraction.
In principle, the above estimates thus provide only a lower limit on the correction needed to determine the significance of such a structure.
However, the negative significance edges suggest that the ring is not much thicker than $0.6\dgr$.

\subsection{Structure in the $S$ distribution}

In order to further quantify the effect that background subtraction has on the signal, we examine the distribution of \gama-ray significance pixels, using the same $\theta=0.2\dgr$ integration beam applied to the map.
The significance histogram (Figure \ref{fig:VER_Hist}), displayed with $\delta \sigma=0.1$ intervals, %\citep[as in][corresponding to the $68\%$ containment per event; \VER]{PerkinsEtAl08},
can be modeled (\VER) as a single Gaussian function, with an adjusted coefficient of determination $\bar{R}^2=0.9956$.

However, this fit to the histogram shows deviations both near the peak and towards the edges.
Of particular interest are the asymmetric deviations for $|S|>1\sigma$ (see figure inset), where most of the signal is found.
These deviations suggest a more complicated underlying distribution.
Indeed, a simple model combining two Gaussian functions provides a much better fit to the data ($\bar{R}^2=0.9967$; note that $\bar{R}$ improves increasingly slowly near $1$).
The best fit is obtained for Gaussian means $\mu_1=-1.23\pm0.21$ and $\mu_2=0.23\pm0.15$, with corresponding standard deviations $0.69\pm0.12$ and $0.91\pm0.08$.
We also model the data as the sum of two Gaussian functions constrained to have the same standard deviation; the resulting best fit is similar to the unconstrained sum: $\mu_1=-0.96\pm 0.06$ and $\mu_2=0.41\pm0.05$, with standard deviation $0.83\pm0.04$ (again with $\bar{R}^2=0.9967$).

Both fits suggest that a signal is present in the map, with pixels that are on average $\mu_2-\mu_1\simeq 1.4\sigma$ above the noise.
Interestingly, a histogram of the pixels found only in the central, $\sim2\dgr$ wide region of the map agrees much better with a single Gaussian \citep{PerkinsEtAl08}.
This indicates that the additional signal is indeed found at radii $\gtrsim 1\dgr$, in agreement with the inferred ring.
Again, this qualitatively agrees with the high significance found using the mock sample.
The above conclusions are not sensitive to the precise pixel binning, and can also be seen from the $\delta\sigma=0.05$ choice of \VER.

\subsection{Cross-correlations}

Additional evidence for the \gama-ray emission arises by cross-correlating the VERITAS map with some potential tracer $q_j$ of the virial shock or of the Galactic foreground.
For simplicity, we use the significance map $s_j$ to represent the VERITAS data for this purpose.
After binning the two maps onto some common grid, we compute the correlation as
\begin{equation}
\mys = \sum_{j=1}^{\pxN} \frac{\left( s_j-\mu_s \right) \left( q_j-\mu_q \right)}{\pxN^2 \sigma_s\sigma_q \pxN_{0.2}^{1/2}} \coma
\end{equation}
where the means $\mu$, standard deviations $\sigma$, and number of pixels $\pxN$, pertain to the region being examined. The factor $\pxN_{0.2}$ accounts for the correlation between pixels within the $\theta=0.2\dgr$ beam, ensuring that the result is approximately independent of the grid resolution.

The significance of the correlation is determined by generating a large ($>\pxN^2$, converged) sample of mock maps with randomly swapped pixels, and computing the corresponding  significance distribution. For observationally-motivated tracers, such as sky maps at a given band, this provides a robust estimate of the significance without requiring any trial factors. Simulated maps that involve some manipulation, such as rotation or resizing beyond those implied by prior information, require trial factors as discussed below.

Consider the effect the ring background subtraction has on such cross-correlations.
In resemblance of the total significance discussed above, any existing correlation diminishes because extended bright regions experience excessive background subtraction.
However, the effect here is more severe, because background regions lying $\leq \thetar$ away from the bright structure are affected as well, appearing excessively faint.
As such ``dark edges'' appear in the background-corrected map but not in the tracers, the correlation decreases further.

Nevertheless, we still find a considerable correlation of the VERITAS signal with a simulated template in \S\ref{sec:Simulation}, significant correlations with radio signals in \S\ref{sec:Radio}, and no correlation with Galactic foreground tracers in \S\ref{sec:Galactic}.

\section{No extended feature in other VERITAS fields}
\label{sec:OtherFields}

As a consistency check, we examine other VERITAS mosaics, similar to that of Coma, in order to test whether some systematics may generate spurious extended features.
Such mosaics are available around the nova in V407 Cygni, the Crab nebula \citep{Aliu12_V407}, and the BL Lacertae object W Com \citep{Acciari09_WComae}.
The mosaics were taken on the same angular scales and with the same $\myw=0.5\dgr$ wobble mode as in Coma, although they are shallower, slightly higher in energy, and at a high zenith.
They also employ a different, reflection background model, but this is taken into account below.
An additional, deep observation of the dwarf spheroidal galaxy Segue 1 \citep{AliuEtAl12} is more similar to Coma, at a low zenith and with the same ring background model, but only a $\sim (3\dgr)^2$ part of the field of view is available.

Unlike the Coma mosaic, none of these observations shows an apparent large scale feature.
To examine this quantitatively, we apply the same search method described in \S\ref{sec:VERITAS} to V407 Cygni and to Crab, using the corresponding reflection background model with $\thetaon=0.1\dgr$.

\begin{figure}[h]
\centerline{
\epsfxsize=5cm \epsfbox{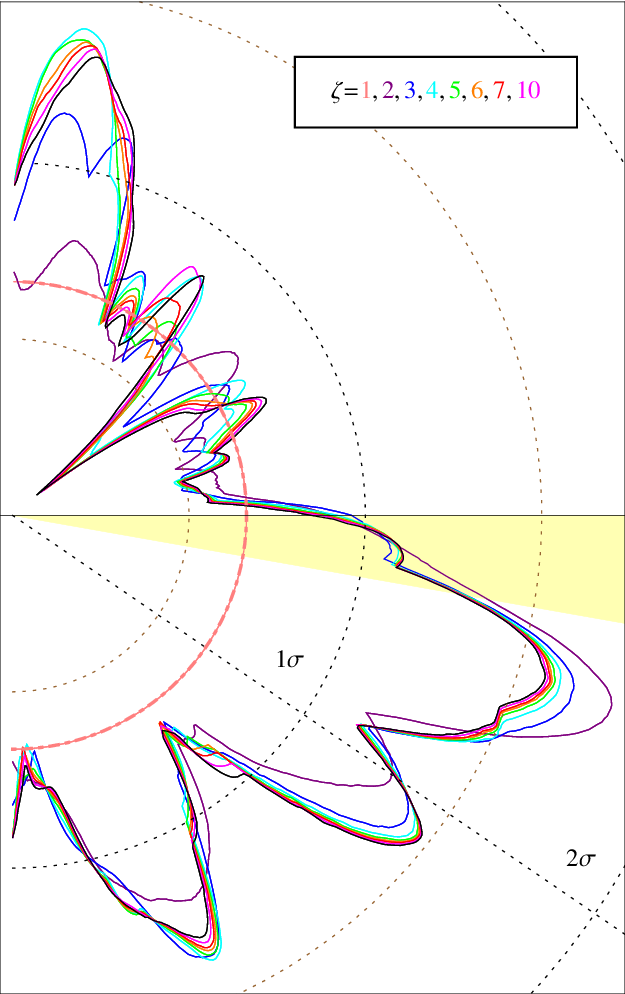}
}
\caption{\label{fig:V407Tilt}
\if \MakeDouble 1
\fontsize{10pt}{8pt}\selectfont
\fi
{\bf Maximal-significance elliptical rings around V407 Cygni.\\}
Notations are the same as in figure \ref{fig:ComaVerTilt}.
\\
}
\end{figure}

In V407 Cygni, the highest significance ring is found with a nominal $S=1.8\sigma$ ($b=1.9\dgr$, $\phi=-18\dgr$).
This is illustrated in Figure \ref{fig:V407Tilt}, which clearly lacks the pronounced directionality seen in Figure \ref{fig:ComaVerTilt}.
(A much lower significance is of course obtained if we use prior information and require, \ApJMark{\eg} $b=\psi_{200}=1.3\dgr$.)
In the reflection model, the true significance $\Sigma$ is much closer to $S$ than in the ring model, because much of the off regions lie outside the extended feature.
This further indicates that the low, $S=1.8\sigma$ nominal significance feature in V407 Cygni is insignificant, even before correcting for trial factors.

\begin{figure}[h]
\centerline{\epsfxsize=8.5cm \epsfbox{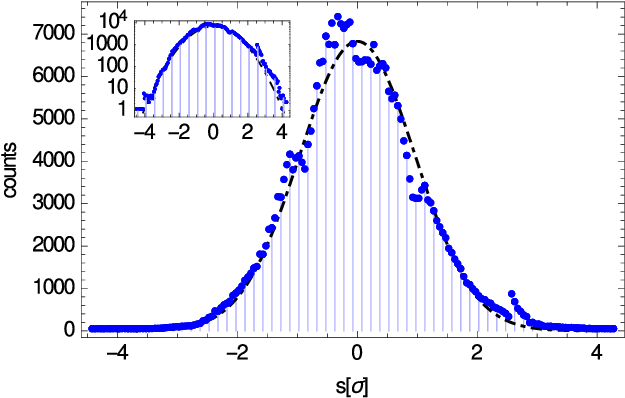}}
\caption{\label{fig:V407_Hist}
\if \MakeDouble 1
\fontsize{10pt}{8pt}\selectfont
\fi
{\bf Pixel significance histogram for V407 Cygni.\\}
A single Gaussian function (dot-dashed) fits the $|S|>1\sigma$ data well.
Notations are the same as in Figure \ref{fig:VER_Hist}.
\\
}
\end{figure}

In Crab, no $S>1\sigma$ ring is found at all, provided that the central $\sim0.2\dgr$ region containing the \gama-ray nebula is excluded.
In both cases, a pixel significance histogram is well-fit by a single Gaussian, with no evidence for an underlying extended source; see for example Figure \ref{fig:V407_Hist}.
Here, including a second Gaussian with $\mu_2>0$ gives a worse fit, lowering the $\bar{R}^2$ value.
To conclude, no extended features are found in VERITAS observations similar to that of Coma, confirming that the {\VER} observation does contain an extended signal in the Coma field.

\section{Gamma-rays from a simulated cluster}
\label{sec:Simulation}

\begin{figure*}[t]
\centering
\subfloat[$\xi_e\mdot=10\%$ \label{fig:SimXi5}]{\epsfxsize=8cm \epsfbox{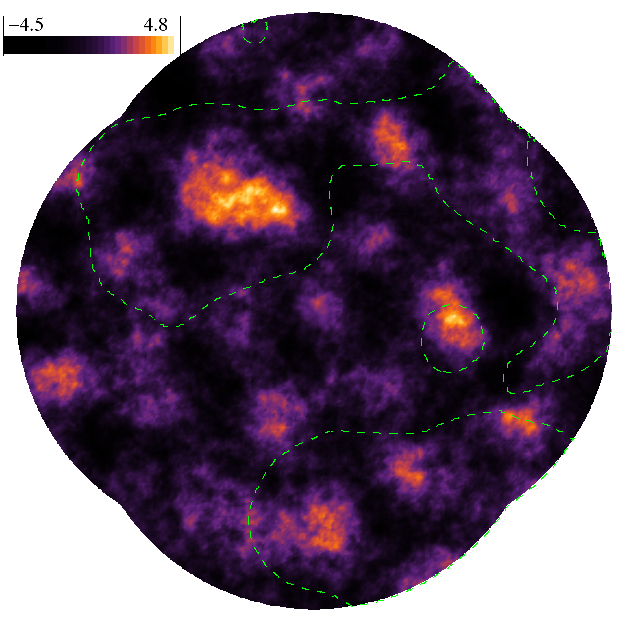}}
\subfloat[$\xi_e\mdot=1\%$ \label{fig:SimXi1}]{\epsfxsize=8cm \epsfbox{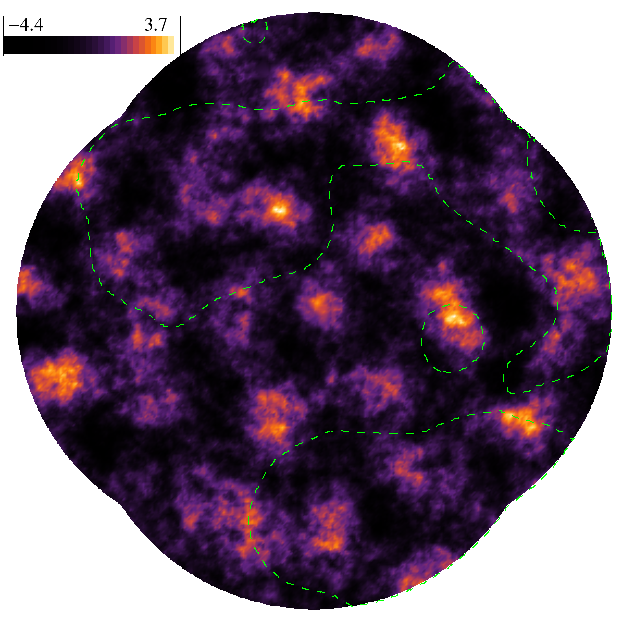}}
\caption{\label{fig:SimCluster}
\if \MakeDouble 1
\fontsize{10pt}{8pt}\selectfont
\fi
{\bf Simulated VERITAS mosaic.\\}
Mosaic of the simulated cluster as would appear in a $220\GeV$ VERITAS observation, using the same $\myw=0.5\dgr$ wobble mode \ApJMark{and ring background} with $\Delta t=18.6$ hour integration.
The simulated ring with bright spots (dashed contours show a threshold on the \gama-ray map of Figure \ref{fig:ComaSim}) is evident for $\xi_e\mdot=10\%$ (left), but can also be seen for $\xi_e\mdot=1\%$ (right) using prior knowledge of the morphology or a template.\\
}
\end{figure*}

The VERITAS signal can be assessed by comparing it to the \gama-ray signature of a simulated, Coma-like cluster in a $\Lambda$CDM simulation \citep{KeshetEtAl03}.
The \gama-ray signal was computed by injecting relativistic electrons at the strong shocks of the simulation, with a Mach number-dependent power-law spectrum given by the Fermi diffusive shock acceleration (DSA) model.
These electrons were assumed to carry a fraction $\xi_e$ of the thermal post-shock energy; see \S\ref{sec:InverseCompton} for details.
We choose the richest cluster in the $(200\Mpc)^3$ simulation box, found \citep{KeshetEtAl04} to have a Coma-like mass $\sim10^{15}M_\odot$ and temperature $k_BT\sim 8\keV$.

\begin{figure}[h]
\centering{\epsfxsize=9cm \epsfbox{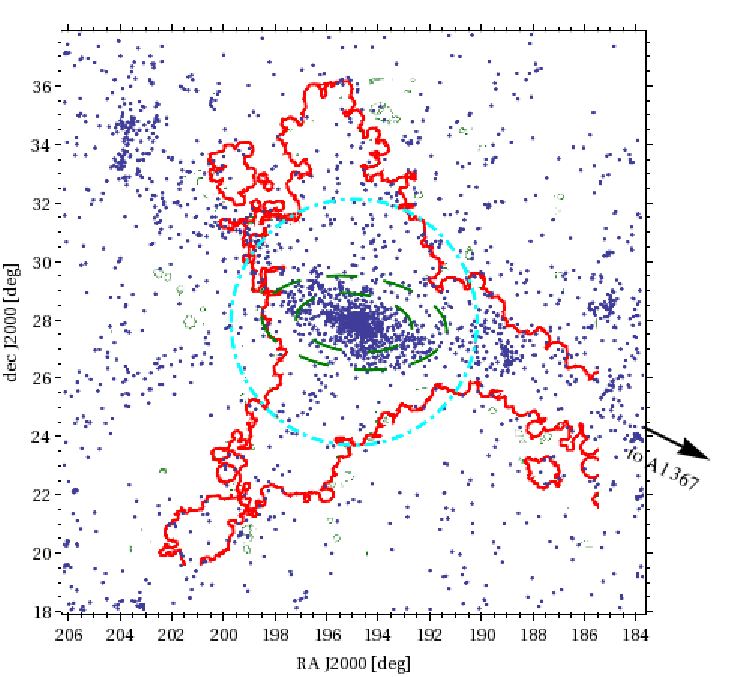}}
\caption{\label{fig:SDSS}
\if \MakeDouble 1
\fontsize{10pt}{8pt}\selectfont
\fi
{\bf The LSS environment of Coma.\\}
SDSS galaxies at redshift $0.018<z<0.028$ near Coma.
The simulation roughly reproduces the filamentary structure near Coma, especially when reflected about the semi-major axis of the virial shock \citep[solid red contours show the density-weighted $k_BT=1\keV$ gas in a $4.4\Mpc$ thick slice through the simulation box; see][]{KeshetEtAl03}.
The dot-dashed cyan circle highlights the region shown in Figure \ref{fig:ComaSim}.
}
\end{figure}

The simulated virial shock (Figure \ref{fig:ComaSim}) has an elliptic structure, elongated towards the main LSS filament (Figure \ref{fig:SDSS}), with semi-minor axis $b\simeq 1.3\dgr$ and semi-major axis $a\simeq 2.5\dgr$ (at the distance of Coma); the average $<220\GeV$ spectral index is $p\simeq 2.05$.
The shock elongation, the morphology of the surrounding filaments, and the enhanced accretion towards the main filament, suggest a prolate geometry aligned with the main filament feeding the cluster.
Unrelaxed, merger clusters such as Coma are indeed known to preferentially be prolate \citep[\eg][]{LemzeEtAl12}.

The center of the simulated cluster is determined by the peak density.
The simulated map is rotated about this point such that the main simulated LSS filament coincides with the $\phi \sim -5\dgr$ orientation of the SDSS filament and the \emph{ROSAT} ellipse.
By coincidence, the agreement between the simulated filaments and the SDSS map remains fairly good even far from the cluster, if the simulated map is inverted along the semi-major axis of the virial shock (Figure \ref{fig:SDSS}).

The simulated \gama-ray signal shows a positive correlation with the VERITAS map (Figure \ref{fig:ComaVerSim}), at the $3.7\sigma$ confidence level.
The correlation remains strong, $>3.5\sigma$ if the maps are rotated by $<5\dgr$ with respect to each other.
The extent of the virial shock may be somewhat overestimated by the adiabatic, smooth particle hydrodynamics (SPH) simulation.
Indeed, the correlation strengthens to $4.1\sigma$ if the simulated map is resized by a factor $0.9$, used henceforth for estimating the model parameters.
However, the enhanced significance beyond $3.7\sigma$ cannot be considered real, as it should be corrected by trial factors.

\begin{figure*}[ht]
\centering
\subfloat[H$\alpha$\label{fig:Halpha}]{\epsfxsize=8cm \epsfbox{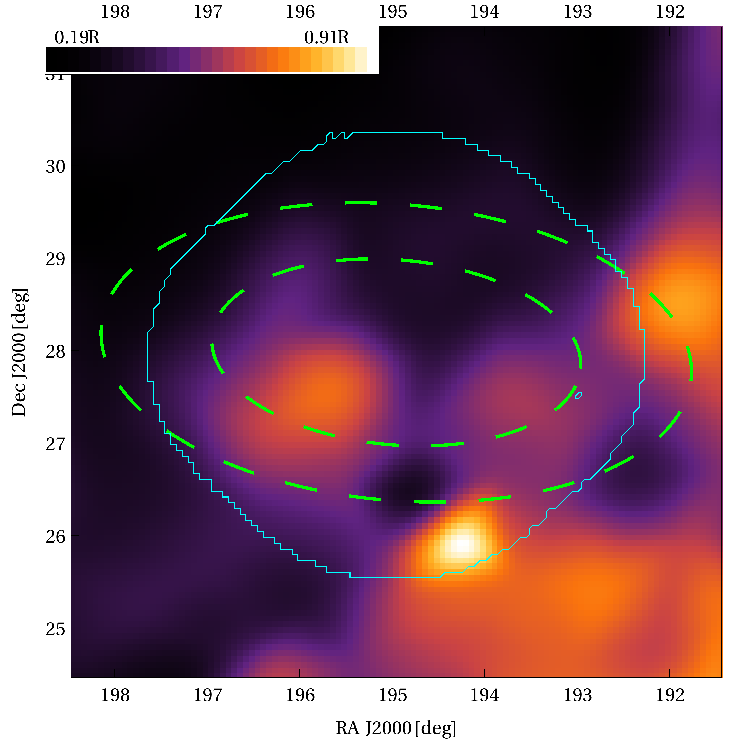}}
\subfloat[Dust\label{fig:SFDdust}]{\epsfxsize=8cm \epsfbox{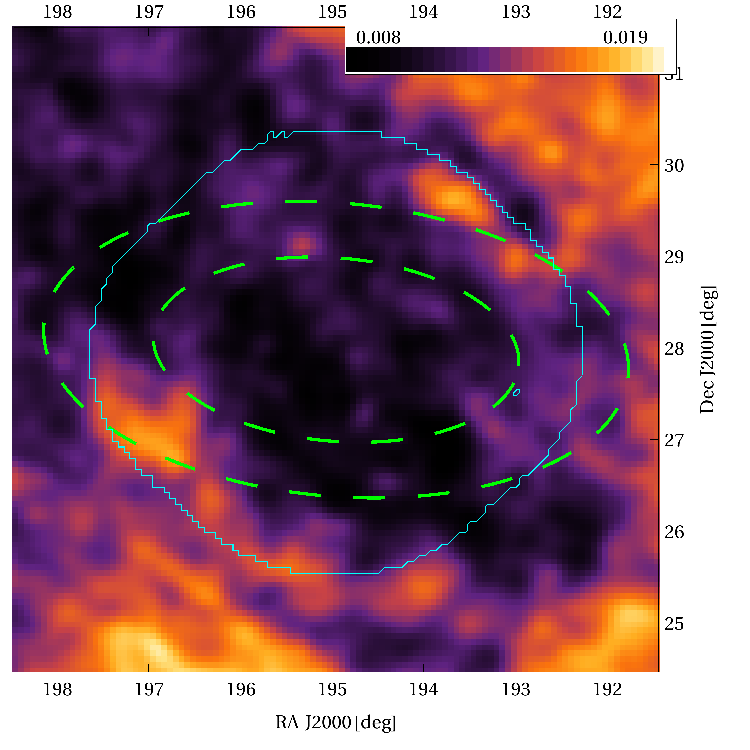}}
\caption{\label{fig:GalacticTracers}
\if \MakeDouble 1
\fontsize{10pt}{8pt}\selectfont
\fi
{\bf Tracers of Galactic foreground near Coma.\\}
The H$\alpha$ full-sky map \citep{Finkbeiner03} (left; colorbar units: Rayleigh) and the SFD dust map \citep{Schlegel98} (right; colorbar: E(V-B) reddening) show only little Galactic structure in the region of the VERITAS mosaic (enclosed by a solid cyan contour), and in particular towards the \gama-ray ring (dashed green contours). \\
}
\end{figure*}

The mass accretion rate $\dot{M}$ of the cluster, normalized by the mass-to-Hubble-time ratio in the dimensionless parameter $\mdot\equiv \dot{M}/(M H)$, varies among clusters and fluctuates in time, and is found \citep{KeshetEtAl04} to be somewhat low when averaged over the simulated cluster, $\mdot\simeq 0.3$.
The \gama-ray map shown in Figure \ref{fig:ComaSim} corresponds to a normalized average accretion rate, $\mdot=1$, and an acceleration efficiency $\xi_e=1\%$.
Equivalently, one may consider the accretion rate of thermal energy $U_{th}$, using the dimensionless parameter $\udot\equiv \dot{U}_{th}/(U_{th} H)$.
In the simulation, $\udot\simeq \mdot$ \citep{KeshetEtAl03,KeshetEtAl04}, although slightly larger, $\udot=(5/3)\mdot$ values arise in a simple isothermal sphere model \citep{WaxmanLoeb00}.

To compare between the VERITAS signal and the expected signature of the simulated cluster, we simulate a $220\GeV$ VERITAS observation with the same parameters as in {\VER} (exposure time $t=18.6$ hours with effective area $A\simeq 4\times 10^8\cm^2$). To the signal we add strong Poisson noise at the VERITAS background level, and filter through the simulated acceptance \citep{MaierEtAl08} of a $\myw=0.5\dgr$ wobble mode. On and off regions are chosen according to the ring model \citep{BergeEtAl07}, and the significance is computed according to the likelihood ratio method \citep{LiMa83}. The results are shown in Figure \ref{fig:SimCluster}, for $\xi_e\mdot=10\%$ and $\xi_e\mdot=1\%$. The signal is seen in both cases, although the latter requires a search specifically targeting the expected signal.

In conclusion, the simulated signal of a virial shock around Coma agrees with the VERITAS ring in morphology, size, and flux, for $\xi_e\mdot$ of order $1\%$.

\section{Negligible Galactic contamination}
\label{sec:Galactic}

When averaged over large angular scales, the Galactic foreground anticipated above $220\GeV$ based on \Fermi observations \citep{AckermannEtAl12} is lower by at least a factor of a few than the ring signal (see \S\ref{sec:InverseCompton}).
Moreover, Galactic signals on $\gtrsim\thetar=0.5\dgr$ scales are efficiently removed from the VERITAS map by the ring background subtraction.
Next, we argue that no significant Galactic contamination on $\lesssim\thetar$ scales is expected in the VERITAS mosaic of Coma.

The main \gama-ray Galactic foregrounds at high latitudes arise from \myNi inverse Compton emission from cosmic-ray (CR) electrons;
\myNii relativistic bremsstrahlung of CR electrons with the interstellar plasma; and
\myNiii nucleon-nucleon scattering between CR ions and the interstellar medium.
The CRs themselves are unlikely to show significant structure on $<0.5\dgr$ scales, as this would require $<10\pc$ confining magnetic fields at $<1\kpc$ distances.
Moreover, synchrotron maps \citep{Haslam82, Reich82} show relatively very little emission from the Coma region.
Therefore, the Galactic foreground on small scales is dominated by electron bremsstrahlung and by $\pi^0$ decay from nucleon collisions, both approximately following the distribution of gas along the line of sight.

\begin{figure*}[ht]
\centering
\subfloat[Thermal SZ\label{fig:WMAP_SZ}]{\epsfxsize=8cm \epsfbox{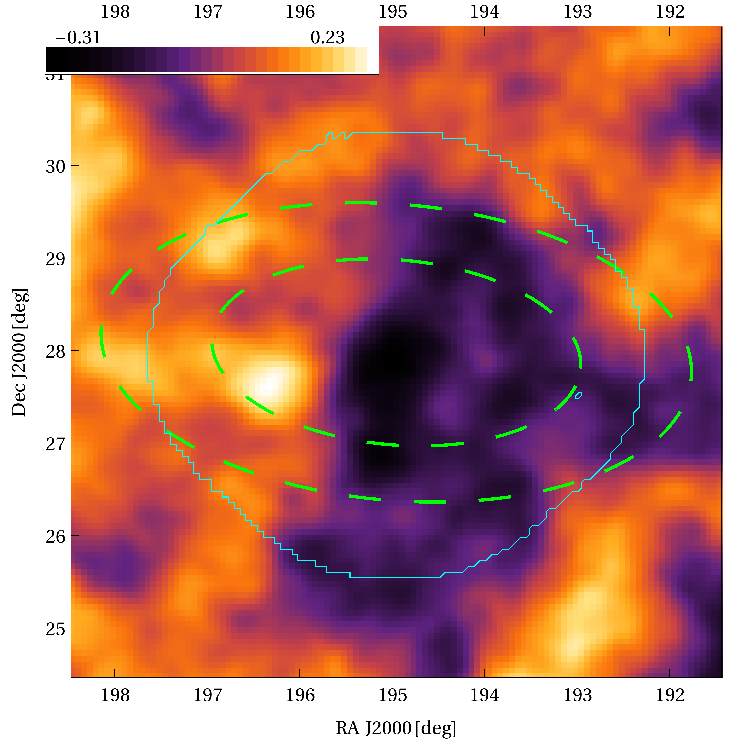}}
\subfloat[Synchrotron\label{fig:WMAP_Syn}]{\epsfxsize=8cm \epsfbox{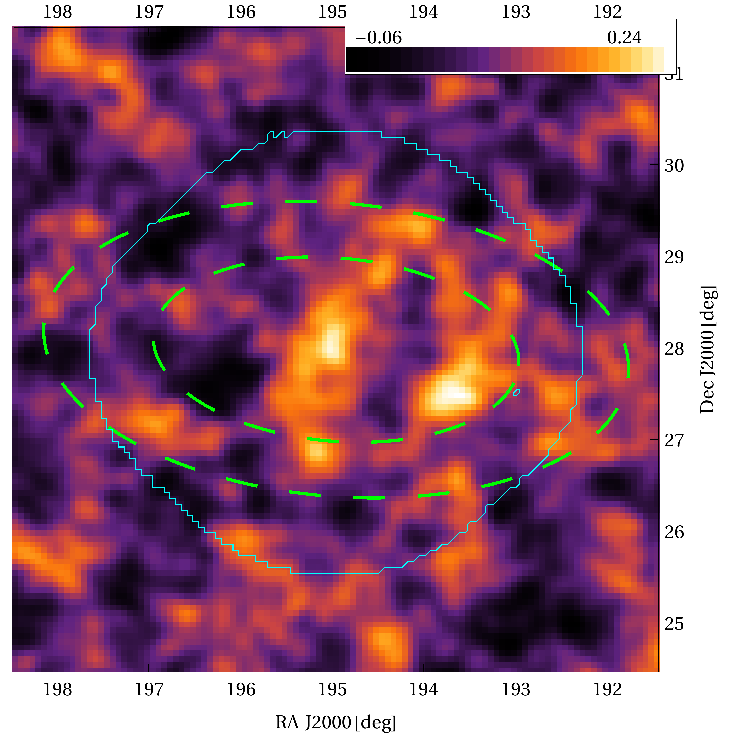}}
\caption{\label{fig:WMAP_Regression}
\if \MakeDouble 1
\fontsize{10pt}{8pt}\selectfont
\fi
{\bf Radio maps of the Coma region.\\}
Thermal SZ (left; colorbar is the low-frequency $\Delta \TSZ=-2y\Tcmb$ in mK) and synchrotron (right; colorbar is $\Tsyn (\nu_0/23\GHz)^{3.2}$ in mK) maps, spectrally extracted from the WMAP seven-year all sky map \citep{KomatsuEtAl11}.
The dashed and solid curves are the same as in Figures \ref{fig:ComaSim} and \ref{fig:GalacticTracers}.\\
}
\end{figure*}

Hence, we may estimate the Galactic contamination remaining in the VERITAS map using gas tracers, such as the full-Sky H$\alpha$ \citep{Finkbeiner03} and infrared dust emission \citep{Schlegel98} maps (Figure \ref{fig:GalacticTracers}).
Both tracers indicate a very low level of Galactic structure in the Coma region, and towards the \gama-ray ring in particular, with only minor structure found mainly in the east.
We confirm this using a spectral linear regression of the seven-year WMAP data, outlined in \S\ref{sec:Radio}.
Both tracers show no positive correlation with the VERITAS map, ruling out any significant remaining Galactic contamination.

Note that synchrotron tracers should not be used here to estimate or eliminate the Galactic foreground.
First, these tracers are inefficient as they follow the smooth CR electron distribution.
More importantly, they are expected \citep{KeshetEtAl04} to \ApJMark{contain} an interesting synchrotron signal from the Coma virial ring itself, which we now pursue.

\section{Significant cross-correlations between \gama-ray and peripheral radio signals}
\label{sec:Radio}

In order to quantify the coincidence between \gama-ray and radio signals, we analyze the WMAP seven-year, full-sky maps \citep{KomatsuEtAl11}.
We use a simple spectral decomposition of the five WMAP bands into synchrotron and SZ maps, avoiding any spatial masking that may complicate the comparison to the VERITAS map.

The synchrotron spectrum is approximated as a pure power-law, with specific brightness $I_\nu\propto \nu^\alpha$ and spectral index $\alpha=-1.2$, typical of cluster radio sources; the results are not sensitive to the precise spectrum for $-2 \lesssim \alpha \lesssim -1$.
The SZ signal is clear in the central parts of Coma, but is difficult to spectrally separate from the CMB fluctuations at large radii \citep{KomatsuEtAl11}.
For simplicity, we leave these CMB fluctuations in the SZ map, thus somewhat diluting any correlations with the VERITAS map.

In general, the WMAP data include non-negligible foregrounds from Galactic synchrotron, dust, and free-free emission \citep{GoldEtAl11}, which are strong towards the Galactic plane and in principle should be removed.
However, as discussed in \S\ref{sec:Galactic}, these signals are particularly weak in the Coma region (Figure \ref{fig:GalacticTracers}), are spatially too smooth (in particular the synchrotron foreground) to have a \gama-ray counterpart surviving the ring background removal, and their tracers do not correlate with the VERITAS map.

Moreover, the extragalactic synchrotron signal cannot be separated from the Galactic $\alpha\simeq -1$ synchrotron, $\alpha\simeq -0.7$ dust, and $\alpha\simeq -0.35$ free-free signals, using spectral regression alone, in particular considering the varying synchrotron spectral index in the cluster radio sources.
Therefore, we leave these faint and probably smooth Galactic contaminations in our synchrotron extragalactic tracer.
This only dilutes the correlations between this tracer and the VERITAS map, as the Galactic signals show no positive correlations with the latter.

Therefore, we fit the five WMAP channels by
\begin{eqnarray}
\Delta T & = & T-\Tcmb \\*
& = & y \Tcmb \left[ \tilde{\nu}\coth\left(\frac{\tilde{\nu}}{2}\right) -4 \right] + \Tsyn \left( \frac{\nu}{\nu_0}\right)^{-3.2} \coma \nonumber
\end{eqnarray}
where $\tilde{\nu}=h\nu/k_B \Tcmb$ is the dimensionless frequency and $y$ is the Comptonization parameter.
The low-frequency SZ temperature shift $\Delta \TSZ=-2y$ and the synchrotron brightness temperature $\Tsyn$ (at the arbitrary frequency $\nu_0$) are used as fit parameters, giving the tracer maps shown in Figure \ref{fig:WMAP_Regression}.

\begin{figure}
\centering
{\epsfxsize=8cm \epsfbox{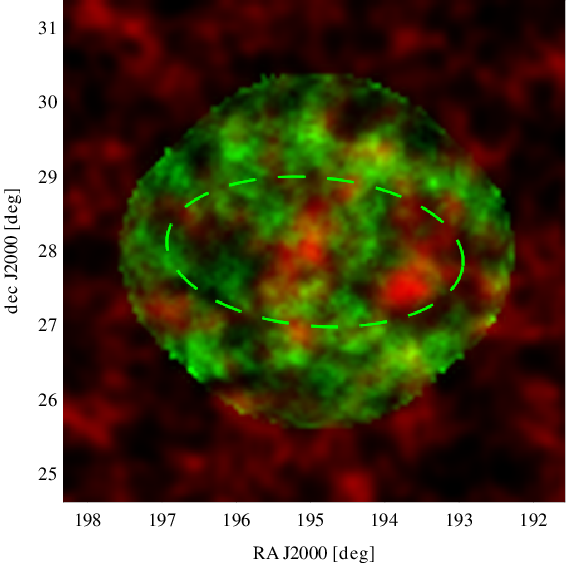}}
\caption{\label{fig:CompositeVERsyn}
{\bf Composite \gama-ray and synchrotron emission.\\}
Superimposed VERITAS map (green; colorbar range: $-3.4\sigma$ to $+3.0\sigma$) and synchrotron map (red; spectrally extracted from the WMAP 7-year full sky-map; colorbar range: $-1.3$ to $+5.4(\nu_0/\mbox{GHz})^{3.2}\K$), showing a positive correlation (yellow) outside the inner edge of the \gama-ray ring (dashed contour) and an anti-correlation (spatially separated green and red, the latter dominated by the radio halo and relic) inside the ring.\\
}
\end{figure}

Next, we compute the cross-correlation amplitudes of the VERITAS data with these $\Tsyn$ \ApJMark{(see Figure \ref{fig:CompositeVERsyn})} and $\Delta \TSZ$ maps.
On average, the VERITAS data show no correlation with the synchrotron map, and an insignificant, $-1.0\sigma$ anti-correlation with the SZ map (\ie a positive correlation with the $y$-parameter).
Note that a \ApJMark{virial} radio signal should generate a positive $\Tsyn$ correlation and a negative $\Delta\TSZ$ correlation.

Features seen in both SZ and synchrotron maps are more pronounced in the western part of the cluster, possibly due to some eastern foreground (\eg see Figure \ref{fig:Halpha}) or CMB fluctuation.
Consequently, for $\mbox{RA}<195.3$ (no more than $0.5\dgr$ east of the cluster's center), the SZ anti-correlation strengthens to $-1.6\sigma$.
Note that this signal does not arise from the spatial separation between the central SZ decrement and the peripheral \gama-ray ring, as this would correspond to a positive correlation here.

More importantly, when we split the data into two regions, outside and inside the inner $b=1\dgr$ ellipse bounding the \gama-ray signal, correlations emerge between \gama-rays and both radio signals.
Outside this inner ellipse, i.e. along the \gama-ray ring, the VERITAS map correlates with the synchrotron map at the $+2.8\sigma$ confidence level, and anti-correlates with the SZ map at the $-2.6\sigma$ level.
The signal is dominated by the western half of the ring, where these correlations strengthen to $+3.2\sigma$ with the synchrotron and $-3.9\sigma$ with the SZ.
The results depend somewhat on the precise region examined.
For example, including the $0.5\dgr$ region east of the center, the synchrotron correlation strengthens to  $+3.5\sigma$ while the SZ correlation remains unchanged.

Within the $b=1\dgr$ ellipse, the VERITAS map does not correlate with the SZ signal, but does show a $-1.9\sigma$ anti-correlation with the synchrotron map, mainly in the western part, where it reaches $-2.2\sigma$.
The inner anti-correlation between the VERITAS and synchrotron maps, evident in the \gama-ray underluminous radio halo and radio relic, and their outer positive correlation, are illustrated in Figure \ref{fig:CompositeVERsyn}.

In the periphery of the cluster, the positive correlation between the VERITAS map and the synchrotron signal agrees with predictions for the coincident inverse-Compton and synchrotron emission from the same relativistic, shock-accelerated electrons, as we show in \S\ref{sec:InverseCompton} and \S\ref{sec:RadioTheory} below.
The anti-correlation with the SZ map agrees with the anticipated SZ signal from the thermal gas, expected to cutoff sharply as the thermal pressure drops beyond the virial shock; see \S\ref{sec:RadioTheory}.

\section{The \gama-ray signal agrees with predictions for a mean \lowercase{$\xi_{e}\dot{m}\sim 1\%$} electron injection rate}
\label{sec:InverseCompton}

Cluster virial shocks last for long, $\sim H^{-1}$, Hubble timescales, and are strong, in particular where cold gas accretes from the voids and the Mach number squared is $\gg 10$.
Therefore, such shocks are thought to accelerate a flat, $dn/d\Ee \propto \Ee^{-2}$ spectrum (constant energy per logarithmic energy interval) of relativistic electrons up to high, cooling-limited energies.
The dominant cooling process here is Compton scattering off CMB photons, implying an electron energy cutoff \citep{LoebWaxman00}
\begin{eqnarray}
\Ee_{max} & \simeq & \frac{\Gamma+1}{2}m_e c \sqrt{\frac{3 e B k_B T}{(\Gamma-1)m_p\sigma_T\ucmb}} \nonumber \\
& \simeq & 60(T_{10}B_{0.1})^{1/2}\TeV \coma
\end{eqnarray}
where $k_B T=10T_{10}\keV$ and $B=0.1B_{0.1}\muG$ are the temperature and magnetic field amplitude downstream of the shock.
Here, $e$ and $m_e$ are the electron charge and mass, $\sigma_T$ is the Thompson cross section, and $c$ is the speed of light.
The photon energy $\Eph\sim 220\GeV$ which maximizes the sensitivity of the {\VER} observation corresponds to electrons with energy $\Ee \simeq (\Eph/3k_B\Tcmb)^{1/2} m_e c^2 \simeq 9\TeV$, below but not far from the estimated cutoff.

The main uncertainty in modelling the nonthermal emission from a virial shock stems from the presently poor understanding of particle acceleration and magnetization in collisionless shocks.
The inverse-Compton emission from the shock can be estimated by assuming that a fraction $\xi_e=0.01\xi_{e,1}$ of the post-shock thermal energy is deposited in relativistic electrons.
Electrons with energies $\gg 100\MeV$ cool quickly, in a narrow shell behind the shock, so the logarithmic emissivity per unit surface area of the shock is
\begin{eqnarray} \label{eq:ICEmissivity}
\Eph^2 \frac{dN_e}{dt\, dA\, d\Eph} & \simeq & \frac{3\xi_e v n k_B T}{4\ln (\Ee_{max}/m_e c^2)} \\
& \simeq & 4.8 \times10^{-8} \xi_{e,1} n_{-4} T_{10}^{3/2} \erg \se^{-1} \cm^{-2} \nonumber \\
& \simeq & 5.1 \times10^{-8} \xi_{e,1} \erg \se^{-1} \cm^{-2} \coma \nonumber
\end{eqnarray}
where $n=10^{-4}n_{-4}\cm^{-3}$ is the downstream (thermal) electron number density,
\begin{equation}
v=\left[\frac{(\Gamma-1)k_B T}{2\mass}\right]^{1/2}=\left(\frac{k_B T}{3\mass}\right)^{1/2}
\end{equation}
is the downstream velocity (with respect to the shock),
and in the last line \ApJMark{of Eq.~(\ref{eq:ICEmissivity})} we used the estimated parameters \ApJMark{(temperature and $R_{200}$ density)} of Coma (see \S\ref{sec:BetaModel}).

The \gama-ray signal from a Coma-like cluster can be computed using Eq.~(\ref{eq:ICEmissivity}) in the framework of a cosmological simulation \citep{KeshetEtAl03, Miniati02}, without introducing additional free parameters.
However, the comparison between simulated and observed clusters is complicated by a secondary uncertainty, involving the local accretion rate of the cluster.
This rate varies spatially along the shock front, fluctuates in time, and differs even among similar clusters as a function of their specific environment.
We parameterize the average accretion rate using the dimensionless variable
\begin{equation}
\mdot \equiv \frac{\dot{M_b}}{M_b H} \simeq \frac{\dot{M}}{M H} \simeq \frac{4\pi R^2 \mass n v}{f_b M H} \coma
\end{equation}
where subscript $b$ refers to baryons, and we assumed a spherical shock of radius $R$ and the cosmic value of $f_b$ at the virial radius.
In the Coma $\beta$-model discussed in Appendix \ref{sec:BetaModel}, $\mdot\simeq 4.3$ (see Eq.~\ref{eq:mdotBeta}), whereas in the simulated cluster \citep{KeshetEtAl04}, $\mdot\simeq 0.3$.
These are averaged quantities, with strong local fluctuations seen in simulations and expected in reality.

Consider the emission from the near vicinity of a cylindrical shock of radius $R=d_A b$, with its symmetry axis perpendicular to the line of sight, where $d_A$ is the angular diameter distance.
A line in the plane of the sky, perpendicular both to this symmetry axis and to the line of sight, cuts the emitting region into a ring of radius $R$ and of negligible thickness.
A small segment of this line, defined by projected distances $\{R_1,R_2\}<R$ from the symmetry axis, corresponds to a ring segment of length $l=2R|\arcsin(R_2/R)-\arcsin(R_1/R)|$.
Consider for simplicity a square beam of solid angle $\vartheta^2=\pi\theta^2$ placed at the edge of the shock, such that $R_1=R$ and $R_2=R-d_A\vartheta$.
Then $l=2d_A b \arccos(1-\vartheta/b)$, and the received (subscript $r$) number flux of $>\Eph$ energy photons becomes
\begin{eqnarray} \label{eq:JperBeam}
J(>\Eph) & \equiv & \frac{dN_r}{dt\, dA\, d\Omega} \\
& \simeq & \frac{b/\vartheta}
{2\pi (1+z)^4} \arccos\left(1-\frac{\vartheta}{b}\right) \Eph \frac{d N_e}{dt\, dA\, d\Eph} \fin \nonumber
\end{eqnarray}

A similar result is expected in the prolate shock inferred in Coma, where combining Eqs.~(\ref{eq:ICEmissivity}) and (\ref{eq:JperBeam}) with the VERITAS map parameters yields
\begin{eqnarray} \label{eq:JPredicted}
J (&>& 220\GeV)  \simeq \frac{\sqrt{b/2\vartheta}} {\pi (1+z)^4} \Eph \frac{d N_e}{dt\, dA\, d\Eph} \coma \\
& \simeq & 5.6 \times 10^{-8} \left( \frac{b}{4\vartheta} \right)^{1/2} \xi_{e,1} n_{-4} T_{10}^{3/2} \se^{-1} \cm^{-2} \sr^{-1} \nonumber \\
& \simeq & 6.0 \times 10^{-8} \left( \frac{b}{4\vartheta} \right)^{1/2} \xi_{e,1} \se^{-1} \cm^{-2} \sr^{-1} \nonumber
\fin
\end{eqnarray}
Here, we used the Coma parameters in the last line.
The approximation made in the first line of Eq.~(\ref{eq:JPredicted}) is good to better than $\sim3\%$ for $\vartheta<b/3$.

To crudely compare this estimate with the \gama-ray structure seen in the VERITAS map, recall that a $1\sigma$ fluctuation corresponds roughly to an excess of $\phN_{\sigma}=\phN_{0.2}^{1/2}\simeq 32$ photons with $\Eph\gtrsim 220\GeV$ in a beam of radius $\theta=0.2\dgr$.
The map was produced by a $t=18.6$ hour exposure (\VER), with an $A\simeq 4\times 10^8\cm^2$ effective area (at $\sim220\GeV$, see \citet{MaierEtAl08}).
A $1\sigma$ signal in the beam therefore corresponds to a photon flux of order
\begin{equation} \label{eq:IVer}
J_{1\sigma} \simeq \frac{\phN_{\sigma}}{\pi \theta^2 t A} \simeq 3 \times 10^{-8} \se^{-1} \cm^{-2} \sr^{-1} \fin
\end{equation}
The mean signal inferred from the histogram (Figure \ref{fig:VER_Hist}), $\mu_2-\mu_1\simeq 1.4\sigma$, is thus comparable to the predicted signal in Eq.~(\ref{eq:JPredicted}) for standard parameters and $\xi_e\sim 0.7\%$.
For a more accurate comparison, we must integrate over the ring and take into account the variable acceptance.

The total photon number flux from the virial shock can be estimated by taking $\vartheta\to 2b$ in Eq.~(\ref{eq:JperBeam}), and integrating over the solid angle \ApJMark{subtended} by the shock.
For a cylindrical shock of radius $b$ and length $2\myzetaT b$, this gives
\begin{eqnarray}
F(>\Eph)  & \equiv & \frac{dN_r}{dt\, dA} \simeq  \frac{b^2\myzetaT}
{(1+z)^4}  \Eph \frac{d N_e}{dt\, dA\, d\Eph} \\
 & \simeq  & 9\times 10^{-11}\frac{\myzetaT b_1^2\xi_{e,1} n_{-4} T_{10}^{3/2} }
{\Eph_{100}(1+z)^4} \se^{-1} \cm^{-2} \nonumber
\coma
\end{eqnarray}
where $b_1\equiv (b/1\dgr)$ and $\Eph_{100}\equiv (\Eph/100\GeV)$.
For the VERITAS band, we use the isothermal $\beta$-model of Coma to obtain
\begin{equation} \label{eq:CylindricalEmission}
F(>220\GeV) \simeq 2.1\times 10^{-10}\myzeta_3\xi_{e,1}\left(\frac{b}{1.3\dgr}\right)^2\se^{-1}\cm^{-2}
\end{equation}
for a $\myzetaT=3\myzeta_3$ cylindrical shock of angular semi-minor axis $b$, or alternatively (see Eq.~\ref{eq:ICFluxBeta})
\begin{equation} \label{eq:SphericalEmission}
F(>220\GeV) \simeq 5.3\times 10^{-11}\xi_{e,1}\psi_{1.3}^2\se^{-1}\cm^{-2}
\end{equation}
for a spherical shock of angular radius $\psi=1.3\dgr\psi_{1.3}$.

Integrating $\non$ and $\noff$, we find that the excess signal in the VERTIAS ring amounts to $(410\pm 150)$ counts, which corresponds to
\begin{equation}
F(>220\GeV)\simeq (1.5\pm0.6)\times 10^{-11}\se^{-1}\cm^{-2}\fin
\end{equation}
Matching this with the cylindrical, $b\simeq1.3\dgr$ signature in Eq.~(\ref{eq:CylindricalEmission}) indicates an acceleration efficiency
\begin{equation}
\xi_e\simeq \myzeta_3^{-1}0.1\% \quad \mbox{and so} \quad \xi_e\mdot\simeq \myzeta_3^{-1}0.3\% \fin
\end{equation}
Using, instead, the spherical shock signature in Eq.~(\ref{eq:SphericalEmission}) corresponds to $\xi_e\simeq \psi_{1.3}^{-2}0.3\%$, and so $\xi_e\mdot\simeq \psi_{1.3}^{-2}1.2\%$.
These estimates are consistent, within our (factor of a few) systematic uncertainties, with the $\xi_e\mdot\sim1\%$ suggested by the simulation in \S\ref{sec:Simulation}.

The above analysis assumed a flat spectrum of accelerated electrons, leading to a flat ($p=2$) photon spectrum.
Indeed, comparing the VERITAS measurement with the EGRET upper limit implies that the photon spectral index between $100\MeV$ and $220\GeV$ cannot be much softer than $p=2.1$.
As the VERITAS measurement probes only photons near $220\GeV$, the above estimates strictly pertain to electrons in energies near $10\TeV$.
Here, we find a logarithmic electron injection \ApJMark{rate}
\begin{equation}
\mdot\Ee \frac{d\xi_e}{d\Ee} \simeq \frac{\xi_e\mdot}{\ln(\Ee_{max}/\Ee_{min})} \simeq 0.05\% \fin
\end{equation}

\section{Synchrotron and SZ signals broadly agree with predictions for $\xi_B\sim 1\%$ magnetization}
\label{sec:RadioTheory}

The same electrons that scatter CMB photons up to the \gama-ray band also emit synchrotron radiation, as they gyrate in the shock-amplified magnetic fields.
The ratio between radio and \gama-ray emission from a strong virial shock is approximately given by the ratio between magnetic and CMB energy densities,
\begin{equation} \label{eq:gamma_to_synch}
\frac{\nu I_\nu}{\Eph J(>\Eph)} \simeq \frac{B^2/8\pi}{\ucmb} \fin
\end{equation}

In order to compute the synchrotron signal, some assumption must be made concerning the magnetic field behind the shock. Typically \citep{WaxmanLoeb00,KeshetEtAl04}, one assumes that a fraction $\xi_B\equiv 0.01\xi_{B,1}$ of the thermal energy is deposited in downstream magnetic fields, based on observations of supernova remnant shocks.
However, this assumption has not been tested until now near a virial shock, and the value of $\xi_B$ is not well constrained.

We may use the coincident \gama-ray and radio measurements to estimate the magnetic field at the shock.
The strongest signal is found in the lowest WMAP frequency, $\nu=23\GHz$, where bright spots along the shock show $\Tsyn\simeq 0.1\mK$, corresponding to a brightness
\begin{equation}
\nu I_\nu = 2\frac{\nu^3}{c^2}k_B \Tsyn \simeq 10^{-9} \erg \se^{-1} \cm^{-2} \sr^{-1} \fin
\end{equation}
This region is part of the VERITAS ring, with $3\sigma$ bright spots that correspond (see Eq. \ref{eq:IVer}) to a \gama-ray brightness $\Eph J(>\Eph)\simeq 3\times 10^{-8} \erg \se^{-1}\cm^{-2} \sr^{-1}$.
Equation (\ref{eq:gamma_to_synch}) then yields $B\simeq 0.6\muG$, which according to the $\beta$-model corresponds to $\xi_B\simeq 1\%$.

We may use the SZ signal near the shock to test the above estimates.
For example, at $\nu=23\GHz$ one expects the ratio between the synchrotron emission and the SZ decrement to be $\Tsyn/\Delta \TSZ \simeq -0.06 \xi_{e,1}\xi_{B,1} \mdot$ for an isothermal sphere distribution \citep{WaxmanLoeb00, KeshetEtAl04}, and $\Tsyn/\Delta \TSZ \simeq -0.4 \xi_{e,1}\xi_{B,1}$ for a $\beta$-model (see Appendix \ref{sec:BetaModel}).

Inspecting the coincident synchrotron ($\Tsyn\sim 0.1\mK$ at $23\GHz$) and SZ ($\Delta\TSZ\sim -0.2\mK$) signals measured along the \gama-ray ring, we find that $\Tsyn/\Delta \TSZ\simeq -0.5$.
This \ApJMark{is consistent with} our estimates $\xi_e\mdot\simeq 1\%$ and $\xi_B\simeq 1\%$ \ApJMark{for the $\beta$-model, but somewhat exceeds the simulated value}.
Note that the parameter $\xi_B$ refers to the magnetic energy weighted over the relativistic electron distribution, and so is elevated by clumping effects \citep{KeshetEtAl04}.

\section{Summary and discussion}
\label{sec:Discussion}

We analyze a $d\sim 4.8\dgr$ diameter VERITAS mosaic (\VER; Figure \ref{fig:ComaVer}) of the Coma cluster, and find evidence for a large-scale, extended \gama-ray feature around the cluster.
It is best described as an elongated ($\myzeta\gtrsim 3$), thick ($\Delta b\gtrsim0.5\dgr$) elliptical ring, with semi-minor axis $b\simeq 1.3\dgr$ (Figure \ref{fig:ComaVerBin}) consistent with the virial radius, oriented approximately along the east-west direction, roughly coincident with the orientation of the LSS filament connecting Coma with Abell 1367 (Figure \ref{fig:SDSS}).

The \gama-ray ring is seen at a nominal $S\simeq2.7\sigma$ confidence level if one adopts the expected signature based on numerical simulations, using the known virial radius, and the expected (east--west) elongation; no trial factors are then necessary.
Varying the model parameters indeed shows that the significance of the signal is maximized at values consistent with the expected (virial) semi-minor axis, with the expected ($\phi\sim -5\dgr$; Figure \ref{fig:ComaVerTilt}) orientation, and with the observationally limited (due to background correction, $\Delta b\sim \thetar+\Delta\thetar\simeq0.6\dgr$) ring width.

The observational wobble mode and ring background model underlying the {\VER} data are not optimized for an extended signal search.
\ApJMark{A control signal analysis,} simulating the effect of this observational pipeline on a large number of mock \gama-ray maps, implies the existence of the ring signal around Coma at a very high, $5.1\sigma$ significance (see \S\ref{sec:TrueSignificance}) even without using any prior knowledge of the size or orientation of the ring. A constrained analysis fixing any of these parameters would yield an even higher significance.

There are several other, independent indications for the presence of an unusual extended signal in the VERITAS mosaic of Coma.
The distribution of pixel significance in the center of the cluster, well inside the extended signal, is consistent with a Gaussian distribution \citep{PerkinsEtAl08}, but this is no longer true when including the region that shows the extended signal (Figure \ref{fig:VER_Hist}).
The VERITAS map of Coma shows the expected correlations with tracers of the shock (see below), and no correlations with Galactic foreground tracers.
The ring-background-subtracted signature of an extended \ApJMark{virial} ring should show a characteristic negative \ApJMark{nominal} significance at its periphery; this is indeed observed.
As a consistency check, we applied our analysis to VERITAS mosaics in other fields (see \S\ref{sec:OtherFields} and Figures \ref{fig:V407Tilt} and \ref{fig:V407_Hist}); none of them shows the signatures of an extended signal as found in Coma.

The size and morphology of the \gama-ray ring agree with predictions for the signature of a virial shock, produced as shock-accelerated electrons inverse Compton-scatter CMB photons.
The VERITAS map thus correlates with a (redshift corrected) simulated \gama-ray map of a Coma-like cluster (Figure \ref{fig:ComaSim}), once the main LSS filaments are similarly oriented (Figure \ref{fig:SDSS}), at the $3.7\sigma$ confidence level with no free parameters.

The brightness of the signal agrees with predictions, provided that a fraction $\xi_{e}\mdot\sim 1\%$ of the thermal energy crossing the virial shock is deposited in relativistic electrons per Hubble time (within a systematic uncertainty factor of a few).
For such efficiencies, mock VERITAS mosaics of the simulated cluster (Figure \ref{fig:SimCluster}) qualitatively match the observed signal.
As a consistency check, we confirm that the brightness, the flux, and the significance histogram yield similar efficiency estimates.

An analysis of the WMAP seven-year data reveals a positive correlation between the \gama-ray ring and synchrotron emission (Figures \ref{fig:WMAP_Syn} and \ref{fig:CompositeVERsyn}), at the $2.8\sigma$ confidence level, and an anti-correlation with the SZ map (Figure \ref{fig:WMAP_SZ}) at $-2.6\sigma$.
The radio signals are clearer in the western half of the ring, towards the LSS filament, where the correlations reach $3.5\sigma$ and $-3.9\sigma$, respectively.
These agree with predictions for the synchrotron emission from the \gama-ray emitting electrons and for the SZ decline with thermal pressure beyond the shock, if the downstream magnetic field is $B\simeq 0.6\muG$, corresponding to $\xi_B\simeq 1\%$ magnetization (within a factor of a few).

The SZ signal we find agrees with the SZ detections by WMAP \citep{KomatsuEtAl11} and Planck \citep{PlanckComa12}.
Note that the \gama-ray ring and these radio features are much farther out than the $\psi\simeq0.5\dgr$ weak shocks reported in X-ray, radio \citep{BrownRudnick11}, and SZ \citep{PlanckComa12} data.

\subsection{Other estimates of \gama-rays from Coma}
\label{subsec:PrevEstimates}

A $3\sigma$ statistical signal from galaxy clusters was previously reported \citep{ScharfMukherjee02}, based on stacking EGRET data around 447 rich clusters.
When compared with source number counts computed analytically \citep{WaxmanLoeb00} and numerically \citep{KeshetEtAl03}, it corresponds to an average $\xi_{e}\mdot\simeq 4\%$, \ApJMark{considerably} higher than we infer in Coma \ApJMark{as it pertains to only the radially projected signal and to low mass ($\sim 10^{13}M_{\odot}$ on average) clusters}.

Subsequent studies have not found such a signal. Instead, upper limits were placed on the emission from galaxy clusters in general, and from Coma in particular, as we next discuss.

Upper limits on the \gama-ray emission from Coma, $F(>100\MeV) <4\times 10^{-8}\se^{-1}\cm^{-2}$, were placed using EGRET data \citep{SreekumarEtAl96, ReimerEtAl03}.
Assuming a flat spectrum, extrapolating \ApJMark{our results} to the EGRET band gives
$F(>100\MeV) \simeq 3\times 10^{-8}\se^{-1}\cm^{-2}$, compatible with the EGRET limit.

An upper limit on the emission from the central region of Coma was imposed using the \Fermi data,
$F(>200\MeV) \lesssim 5\times 10^{-9}\se^{-1}\cm^{-2}$ \citep{AckermannEtAl10}.
This was obtained using an aperture optimized for emission from the central, X-ray bright region of the cluster, rather than from a peripheral, virial ring.
Upper limits on extended emission from Coma,
$F(>500\MeV) \lesssim (4-7)\times 10^{-10}\se^{-1}\cm^{-2}$ \citep[model-dependent range; see][]{AckermannEtAl14_GammaRayLimits}, and specifically on emission from a ring structure around it, $F(>100\MeV) \lesssim 3\times 10^{-9}\se^{-1}\cm^{-2}$ \citep{ZandanelAndo14, Prokhorov14}, were subsequently reported.

However, such limits are complicated by the poor resolution and strong background at these energies. The \Fermi resolution at $100\MeV$ is $\sim5\dgr$, gradually \ApJMark{improving} to $\sim1\dgr$ at $1\GeV$ \citep{AckermannEtAl10, FermiCalibration12}.
The foreground in the northern Galactic cap region is $\sim 2.5\times 10^{-3}\MeV\se^{-1}\cm^{-2}\sr^{-1}$ above $200\MeV$ \citep{AckermannEtAl12}.
This implies a \ApJMark{foreground} flux $F(>200\MeV)\simeq 6\times 10^{-8} \se^{-1}\cm^{-2}$, about $40$ times stronger than the aforementioned limits.
This foreground is neither spatially flat (\eg a bright \gama-ray source lies $3.4\dgr$ from the cluster center \citep{ZandanelAndo14}), nor well constrained \citep[\eg][]{KeshetEtAl04_EGRB}.

In the absence of a well-controlled foreground model, a resolvable template, or a distinctive spectral feature, the systematic uncertainties in the above limits are substantial.
Beating them down to the level of the anticipated signal, \ie a few percent of the background, is challenging.
Indeed, the high resolution and controlled background of VERITAS are crucial for the signal detection.

Several upper limits were imposed by stacking galaxy clusters, using EGRET \citep{ReimerEtAl03} or \Fermi \citep{AckermannEtAl10, HuberEtAl13, AckermannEtAl14_GammaRayLimits, ProkhorovChurazov14, GriffinEtAl14} data, typically focusing on the central, X-ray bright regions.
Of these, the latter, a $2.3\times 10^{-11}\se^{-1}\cm^{-2}$ upper limit at ($0.8$--$100$) GeV energies, is the most restrictive and most relevant for our purpose, because the clusters were stacked on the same physical, rather than angular, scale \citep{GriffinEtAl14}.

Taking into account the average redshift $0.08$ of the 78 rich clusters in the \cite{GriffinEtAl14} sample, their upper limit is about an order of magnitude below what one would expect if all clusters in the sample have Coma-like virial rings.
However,
\myNi while using a large number of clusters effectively reduces the background, it remains high, in particular considering the lower latitudes of most clusters in the sample;
\myNii while the resolution at $0.8\GeV$ is better than at lower energies, the distant clusters subtend smaller solid angles so their virial rings cannot be resolved;
\myNiii the limit pertains to the central $2\Mpc$ \ApJMark{radius} of each cluster, inward of \ApJMark{extended, elongated} virial \ApJMark{shocks as in Coma};
and
\myNiv virial rings are irregular and intermittent (on cosmological timescales); it would be difficult to distinguish them from the background if they form extended filaments.
In summary, the above \Fermi stacking analyses lack the sensitivity and resolution needed to rule out a Coma-like signal from clusters.

\subsection{Implications and outlook}

Inverse-Compton emission from virial shocks can explain the hard X-ray signals observed in several clusters \citep{KushnirWaxman10}, provided that $\xi_e\lesssim 10\%$.
For example, $\xi_e\mdot\simeq 4\%$ was recently inferred from hard X-ray emission from a LSS filament \citep{MakiyaEtAl12}.
The VERITAS ring corresponds to a $\sim 3\eV\se^{-1}\cm^{-2}$ flux at $20\till 80\keV$ energies (assuming a flat spectrum), accounting for part of the $(8.1\pm2.5)\eV\se^{-1}\cm^{-2}$ hard X-ray signal observed \citep{FuscoFemianoEtAl11}.
\ApJMark{Future} high resolution observations in hard X-rays could resolve the virial ring in Coma, and detect virial shocks in many other clusters.

A VERITAS observation deeper than the present $18.6$ hour exposure could decisively test our results, in particular if redundant background removal can be minimized.
This can be achieved by avoiding on and off regions simultaneously falling within the expected extended feature, for example using a reflection model with a minimal separation in declination (for the case of Coma).
This would allow for a more precise reconstruction of the virial shock structure and a determination of variations in the electron deposition rate along the shock front.
Such data would be valuable in the study of large-scale structure formation at low redshift, mapping the feeding pattern of Coma, probing the surrounding voids, the Coma--Abell 1367 filament, and other LSS filaments, and tracing the warm-hot intergalactic medium (WHIM) immediately behind the shock.

Observations at lower energies, for example using the $\sim 50\GeV$ MAGIC \v{C}erenkov telescopes \citep{DeLotto12}, should be able to utilize the higher photon flux to clearly identify the \gama-ray ring.
At yet lower energies, the \Fermi \gama-ray space telescope is marginally sensitive to the signal, but it would be challenging to resolve the shock structure with the \ApJMark{extended} point spread function \citep{BurnettEtAl09}.

The VERITAS energy threshold, $\Eph\simeq 220\GeV$, is not far below the expected photon cutoff, $\Eph_{max}\simeq 10 T_{10}B_{0.1}\TeV$.
Observations at higher energies would become increasingly sensitive to this cutoff, thus probing the conditions at the shock and the nature of particle acceleration.
%, although pair production off the infrared background also becomes important at such energies \citep{FranceschiniEtAl08}.
Interestingly, no evidence for extended emission was found in a $1.1\TeV$ observation of Coma by the HESS telescopes \citep{AharonianEtAl09}.
Despite the higher energy threshold and the shorter, $8.2$ hour observation, it was only $\sim 2.3$ times less sensitive than the VERITAS observation (assuming a flat spectrum), thanks to the larger effective area \citep{BenbowEtAl05}, $\sim 2\times 10^9\cm^2$.
The HESS map shows no significant signal or correlation with the VERITAS data, but the \gama-ray ring may well be shallowly buried under the noise.
Note that the optical depth of the HESS photons due to pair production off the infrared background, estimated \citep{FranceschiniEtAl08} as $\tau\simeq 0.2$, may not be negligible.

Finally, consider the contribution of cluster virial shocks to the extragalactic \gama-ray and radio backgrounds.
These components were computed analytically \citep{LoebWaxman00, TotaniKitayama00, KeshetEtAl03}, and calibrated numerically using cosmological simulations \citep{KeshetEtAl03, Miniati02, KeshetEtAl04}.
While the main model parameters $\xi_e\mdot$ and $\xi_B$ are likely to fluctuate within clusters and vary among different clusters, it is instructive to adopt the values inferred in Coma as typical.
%An accretion rate $\mdot\simeq 1$ is intermediate between the analytic model \citep{WaxmanLoeb00} and the low redshift result of numerical simulations \citep{KeshetEtAl03}.
%Normalizing all clusters by this value, we obtain
This gives a diffuse \gama-ray component $\Eph^2 dJ/d\Eph\simeq 0.1(\xi_{e,1}\mdot)\keV\se^{-1}\cm^{-2}\sr^{-1}$, significantly contributing to the extragalactic background \citep{KeshetEtAl04_EGRB}.
In the radio, the revised parametrization yields a $\nu I_\nu \sim 10^{-11}(\xi_{e,1}\mdot) (\xi_B/1\%)\erg\se^{-1}\cm^{-2}\sr^{-1}$ synchrotron signal, dominating the extragalactic low frequency radio background \citep{KeshetEtAl04, KeshetEtAl04_SKA}.
It should be observable through $\delta T_l\simeq 1 (\nu/\mbox{GHz})^{-3}\K$ fluctuations at multipoles $400\lesssim l \lesssim 2000$ with present interferometers such as LOFAR and EVLA.

\acknowledgements
We thank I. Reiss, I. Gurwich, B. Katz, M. Pohl, M. Ostrowski, F. Zandanel, R. Mukherjee, and L. Rudnick for useful discussions.
The research of UK has received funding from the European Union Seventh Framework Programme (FP7/2007-2013) under grant agreement n\textordmasculine ~PCIG09-GA-2011-293975 and from the IAEC-UPBC joint research foundation (grant No. 257), and was supported by
the \ApJMark{Israel Science Foundation (grant No. 1769/15)} and by the GIF (grant I-1362-303.7/2016).

\onecolumngrid
\appendix

\if \MakeDouble 1
\myonecolumn
\fi

\section{$\beta$ model for the shock}
\label{sec:BetaModel}

An analytic model for emission from the virial shock requires some assumptions specifying the gas distribution.
Simple choices include an isothermal sphere \citep{WaxmanLoeb00} or an isothermal $\beta$-model \citep{KushnirWaxman09}.
We begin with the latter, but as $\beta\simeq 2/3$ in Coma, we then effectively revert to the former.

In our simple model, it is assumed that \myNi the electron number density is approximately given by
\begin{equation}
n_e(\vect{r}) = n_{e,0} \left( 1+\frac{r^2}{r_c^2}\right)^{-\frac{3}{2}\beta}
\end{equation}
out to the virial radius; and \myNii on average, the gas distribution is approximately isothermal and in hydrostatic equilibrium.
In Coma \ApJMark{\citep{FukazawaEtAl04, ChenEtAl07}}, the central \ApJMark{electron} density is estimated as $n_{e,0}=(3.5\pm 0.7)\times 10^{-3} \cm^{-3}$, the core radius as $r_c=343_{-20}^{+22}\kpc$, the profile index as $\beta=0.654_{-0.021}^{+0.019}$, and the (emission measure-weighted) temperature as $k_BT=8.38\pm0.34\keV$.

With the above assumptions, the region enclosing $\mykappa=200\mykappa_{200}$ times the critical mass density $\rho_c$ \ApJMark{of the Universe} is sufficiently large to neglect the core (effectively reverting to the isothermal model, in particular if approximating $\beta=2/3$), and so is characterized by \ApJMark{a radius}
\begin{equation}
R_\mykappa \simeq \left(\frac{9\beta k_B T}{4\pi G \mykappa \mass \rho_c} \right)^{1/2} \simeq 2.3 \mykappa_{200}^{-1/2} \Mpc \coma
\end{equation}
\ApJMark{an enclosed mass}
\begin{equation}
M_\mykappa \simeq \frac{3\beta k_B T r_\mykappa}{G\mass} \simeq 1.4\times 10^{15} \mykappa_{200}^{-1/2} M_\odot \coma
\end{equation}
and \ApJMark{a particle number density}
\begin{equation}
n_\mykappa \equiv \frac{3+5\chi}{2+2\chi} n_e(r_\mykappa) \simeq 1.4 \times 10^{-4} \mykappa_{200} \cm^{-3} \fin
\end{equation}

\ApJMark{The normalized accretion rate through the shock may be computed as}
\begin{equation} \label{eq:mdotBeta}
\mdot_\mykappa = \frac{4\pi R_\mykappa^2 \mass n_\mykappa v_\mykappa}{f_b M_\mykappa H} \simeq
5.7 \mykappa_{200}^{1/2} \, .
\end{equation}
\ApJMark{However, it is preferable to derive $\dot{m}$ in a method less sensitive to the extrapolation of the central density out to the virial radius.
By integrating the baryon mass out to $R_k$ and assuming that it constitutes a fraction $f_b$ of the mass, we find}
\begin{equation}
\dot{m}_\mykappa \simeq \frac{2(1-\beta)}{H} \sqrt{\frac{\pi \mykappa G\rho_c}{3\beta}} \simeq 4.3 \mykappa_{200}^{1/2} \fin
\end{equation}

\ApJMark{We may now compute} the inverse-Compton emissivity per unit shock area,
\begin{eqnarray} \label{eq:ICEmissivityBeta}
\Eph^2 \frac{dN_e}{dt\, dA\,  d\Eph} & \simeq & \frac{3\xi_e f_b \mdot_\mykappa M_\mykappa Hk_B T}{16\pi R_\mykappa^2 \mass \ln (\Ee_{max}/m_e c^2)}   \\
& \simeq & 9.2 \times 10^{-9} \xi_{e,1}\mdot_\mykappa \mykappa_{200}^{1/2} \erg \se^{-1} \cm^{-2} \nonumber \\
& \simeq & 3.9 \times 10^{-8} \xi_{e,1}\mykappa_{200} \erg \se^{-1} \cm^{-2} \nonumber \coma
\end{eqnarray}
the resulting photon flux from a spherical shock,
\begin{eqnarray} \label{eq:ICFluxBeta}
F(>\Eph) & = & \frac{4\pi r_\mykappa^2}{4\pi d_L^2} \Eph \frac{dN_e}{dt\, dA\,  d\Eph} =  \frac{\psi_\mykappa^2}{(1+z)^4} \Eph \frac{d N_e}{dt\, dA\,  d\Eph}   \\
& \simeq & 1.2\times 10^{-11} \frac{\xi_{e,1}\mdot_\mykappa \mykappa_{200}^{1/2}}{(1+z)^4} \left(\frac{\psi}{1.3\dgr}\right)^{2}  \left(\frac{\Eph}{220\GeV}\right)^{-1} \se^{-1} \cm^{-2} \nonumber \\
& \simeq & 5.3\times 10^{-11} \frac{\xi_{e,1} \mykappa_{200}}{(1+z)^4} \left(\frac{\psi}{1.3\dgr}\right)^{2}  \left(\frac{\Eph}{220\GeV}\right)^{-1} \se^{-1} \cm^{-2} \coma \nonumber
\end{eqnarray}
where $\psi$ is the angular radius of the shock,
the brightness in a square $\vartheta^2$ beam at the edge of the shock in the cylindrical shock limit,
\begin{eqnarray} \label{eq:ICBrightnessBeta}
J(>\Eph) & \simeq & \frac{\sqrt{b/2\vartheta}}{\pi(1+z)^4} \Eph \frac{dN_e}{dt\, dA\,  d\Eph}
=\frac{15f_b H \xi_e\mdot}{4(1+z)^4\epsilon \log\gamma_{max}} \left( \frac{\beta b \mykappa \rho_c}{G\vartheta} \right)^{1/2} \left( \frac{k_B T}{\pi \mass} \right)^{3/2}\\
& \simeq & 1.1\times 10^{-8} \xi_{e,1}\mdot_\mykappa \left(\frac{\mykappa_{200}b}{4\vartheta} \right)^{1/2}  \se^{-1} \cm^{-2} \sr^{-1} \nonumber \\
& \simeq & 4.7\times 10^{-8} \xi_{e,1} \mykappa_{200} \left(\frac{b}{4\vartheta} \right)^{1/2}  \se^{-1} \cm^{-2} \sr^{-1} \nonumber \coma \nonumber
\end{eqnarray}
and the magnetic field at the shock,
\begin{equation}
B_\mykappa = \sqrt{\xi_B 8\pi \frac{3}{2} n k_B T} \simeq 0.8 \mykappa_{200}^{1/2}\xi_{B,1}^{1/2} \muG\fin
\end{equation}

%\if \MakeDouble 1
%\myonecolumn
%\fi

%\bibliography{Virial}

\end{document}